# Data-driven and distributed governance of building facilities management using decentralized autonomous organization, digital twin, and large language models

Reachsak Ly, Ph.D.[1] Alireza Shojaei, Ph.D.[2], Xinghua Gao, Ph.D.[3], Philip Agee, Ph.D.[4], Abiola Akanmu, Ph.D.[5]

[1] School of Technology, Eastern Illinois University, Charleston, Illinois, United States
[2,3,4,5] Myers-Lawson School of Construction, Virginia Polytechnic Institute and State University, Blacksburg, Virginia, United States

**Abstract**

While traditional AI and data-driven facilities management approaches have improved building operational efficiency, they remain constrained by centralized organizational structures which are vulnerable to cyber-attacks, limited contextual understanding, and decision-making structures that exclude key stakeholders from the governance process. This paper introduces a novel AI and data-driven distributed governance framework for smart building management that integrates decentralized autonomous organizations (DAOs), digital twins, large language models (LLMs), and blockchain technology. This framework aims to enable transparent collective decision-making through a DAO governance platform, implements data-driven management using IoT and digital twins, incorporates LLM-based virtual assistants for enhanced decision support, and utilizes blockchain for secure building automation. A full-stack decentralized application is developed to facilitate user interaction with these integrated components. The system was evaluated for cost-efficiency, scalability, data security, and usability using the System Usability Scale (SUS). Expert interviews were also conducted to evaluate its practical benefits and challenges.

Keywords: Decentralized autonomous organization, Large language models, Decision-support system, Facility management, Digital twins

## 1. Introduction

Facilities management refers to the practice of coordinating and overseeing the maintenance, operations, and services within a built environment to ensure functional, safe, and efficient working conditions(Nielsen et al. 2016).FM combines the key elements of people, place, processes, and technology with the built environment to enhance quality of life and increase operational efficiency (IFMA). As the built environment incorporates more sophisticated systems and technologies and generates vast amounts of information, data-driven approaches have become significantly important in the field of facilities management. Data-driven facilities management is the practice of using data and analytics to optimize and improve the operations, maintenance, and management of facilities and their assets. It leverages advanced technologies such as the Internet of Things, sensors, building automation systems (BAS), digital twins, artificial intelligence (AI), and data analytics to collect, analyze, and act upon vast amounts of data generated by the physical infrastructure and components. For instance, researchers have also utilized AI techniques such as deep learning (DL), deep reinforcement learning (DRL), and long short-term memory (LSTM) networks to predict equipment failures, optimize energy usage, and improve occupant comfort. In addition, digital twins also provide a dynamic and interactive platform for monitoring building performance, conducting predictive maintenance, and optimizing resource management.

However, despite these advancements, several limitations remain. For instance, conventional building automation systems are often based on centralized protocols, making them vulnerable to risks such as cyber-attacks. According to a comprehensive examination of smart building security conducted by Kaspersky in 2019, nearly four in ten (37.8%) smart building automation systems have been affected by malicious cyber-attacks (Kaspersky 2021). There is also a lack of research exploring decentralized frameworks that combine digital twins with secure, decentralized building automation. Therefore, the research in enhancing the security and resilience of building automation operations is crucial in addressing the vulnerabilities and inefficiencies of the current smart building systems.

Additionally, while traditional AI excels at specific facilities management tasks, it is still trained on predefined rules or specific training data and may lack contextual understanding. Therefore, the capability for a more open-ended analysis and decision support in complex facility management scenarios is still an open research gap. The rise of large language models (LLMs) and their ability to understand and generate human-like responses present a potential opportunity to enhance data-driven FM. LLMs could enable more natural language interaction for querying and analyzing digital twin data and provide decision support in complex facility management scenarios.

Furthermore, the traditional FM operations within the built environment often rely on centralized organizational structures. This approach, which concentrates decision-making authority among a select few, typically senior facility managers, was originally designed to streamline processes and ensure consistency (Xu et al. 2020). However, recent studies have revealed significant drawbacks to this model, particularly in terms of transparency and alignment with the diverse needs of building occupants. The growing dissatisfaction among building users stems from their limited involvement in FM-related decisions (Leaman and Bordass 2001). This lack of participation often results in decisions that fail to adequately address their evolving needs and preferences within their living environment. Moreover, a growing body of research has also highlighted the limitations of this conventional, centralized FM approach and emphasized the importance of community-based facility management (Alexander and Brown 2006; Michell 2010; Adewunmi et al. 2023), a participatory approach that fosters democratized and socially inclusive facility management practices that prioritize the diverse needs and perspectives of related stakeholders (Tammo and Nelson 2014). This approach not only enhances the transparency and accountability of the decision-making process but also encourages the engagement of all community members in shaping the management and operation of the shared facility (Tammo and Nelson 2012).

However, despite decentralizing the decision-making process, the CbFM framework's current coordination mechanisms and incentivization systems still rely on centralized, Web 2.0 technologies. This reliance can compromise trust and efficiency, as it depends on conventional communication and record-keeping methods that lack inherent transparency and security. The absence of a secure, decentralized system can result in a lack of trust, transparency, and accountability among stakeholders regarding decision-making and resource allocation. The advent of decentralized autonomous organizations (DAOs) (Wang et al. 2019a) offers a potential solution to these challenges in traditional community-based facility management. DAOs operate as digital, community-driven entities on blockchain networks, functioning transparently and autonomously with democratic and collective decision-making capabilities among their members. The fundamental operations of DAOs are governed by rules encoded in smart contracts (Singh and Kim 2019), which can enhance transparency, security, and trust in the decision-making processes of CbFM. A smart contract is a self-executing digital contract stored on a blockchain, where the terms are written in code and automatically executed when predefined conditions are met (Naderi et al. 2023).

The integration of AI and a data-driven approach with a decentralized governance mechanism can revolutionize the current facilities management practice. By merging the intelligent insights provided by LLM and data analytics with the democratic and transparent decision-making enabled by decentralized technologies, this integrative framework can provide both the efficient and intelligent as well as the socially inclusive and decentralized FM. However, the current literature has yet to explore the integration between the LLM within the data-driven governance workflow, as well as the synergies between decentralized technologies such as blockchain and DAO with large language models.

To address this knowledge gap, this paper presents novel, AI and data-driven distributed governance frameworks for facilities management using LLM, digital twin, blockchain, and decentralized autonomous organizations. The specific research objectives of this study are the following:

(1) To examine how decentralized autonomous organizations can enable transparent and collective decision-making in the governance of building infrastructure with the development of a decentralized governance platform for facility management.
(2) To implement the data-driven governance framework for the facilities management using digital twin and Internet of things.
(3) Assess the scope and capacity of AI conversational agents in assisting the building infrastructure governance process by integrating the LLM-based virtual assistant into the proposed data-driven and decentralized governance workflow.
(4) Implement a blockchain-based decentralized automation framework for building operations.
(5) To develop a full-stack decentralized application (DApp) to facilitate user interaction with a decentralized governance platform, digital twin visualization, and the LLM-based AI agent.

The remainder of this paper is structured as follows: Section 2 presents the current practices in facilities management, motivation, and challenges of community-based facilities management, followed by an introduction to the relevant concepts of DAOs and blockchain technology and why they are suitable to address the gaps. This section also provides the background on the application of digital twin and data-driven governance in facilities management as well as the application of large language models and DAOs in the construction industry. The research method of the study is presented in section 3 while section 4 describes the proposed AI and data-driven distributed governance FM framework. Section 5 provides the implementation and prototype of the proposed system. Section 6 describes the evaluation and validation of the system. Then discussion of the findings, implications, limitations of the research, and future research directions is made in section 7. Finally, the conclusion is presented in section 8.

## 2. Point of departure

### 2.1. Toward inclusive and decentralized governance for facilities management

Facilities Management (FM) is recognized as a critical process within the building infrastructure that oversees the built assets, personnel, systems, and support services to ensure their alignment with core business objectives (Chotipanich 2004). The scope of FM has expanded significantly in recent years, encompassing a diverse array of services and processes crucial for the efficient operation of buildings and infrastructure which play vital roles in their operational efficiency, safety, and sustainability (Okoro 2023). Traditionally, FM has been characterized by a centralized approach, with decision-making authority concentrated among facilities managers or management teams. The absence of diverse perspectives from building occupants, tenants, and the surrounding community can result in decisions that diverge from the priorities and preferences of those who interact with the building and living environments (Leung et al. 2012). Consequently, this can lead to suboptimal outcomes, reduced occupant satisfaction, and potential conflicts among stakeholders, ultimately impacting. This limitation also hinders the ability to gather valuable insights and knowledge from those directly impacted by FM decisions, ultimately leading to inefficiencies, and missed opportunities for improvement.

In response to these challenges, Alexander and Brown (Alexander and Brown 2006) have introduced a new paradigm in FM, Community-based Facility Management (CbFM), which seeks to develop a more socially inclusive, and participatory model of FM. CbFM is rooted in the principle of managing facilities and services in a manner that reflects the needs and values of the community in the built environment. CbFM recognizes that building occupants, tenants, and community members can contribute their valuable insights and knowledge to improve the built environment's functionality, efficiency, and overall user experience. By actively incorporating these stakeholders in the decision-making process, CbFM aims to cultivate a sense of shared ownership, enhance occupant satisfaction, and promote sustainable practices within the built environment. At its core, CbFM is driven by the desire to create a more user-centric and responsive approach to FM, ensuring that the management and maintenance of the built environment are aligned with the diverse needs and preferences of its occupants.

However, despite the advantages provided by community-based facility management, there are still a few challenges and limitations that hinder its full potential for effective implementation. One of the primary hurdles facing CbFM is the paradoxical reliance on centralized Web 2.0 technologies for coordination, despite its aim to distribute and decentralize decision-making power. This centralization introduces vulnerabilities in terms of trust and operational efficiency, as it depends on traditional communication and record-keeping methods that lack inherent transparency and robust security measures. For example, research by Sedhom et al. (2023) highlights two major challenges in CbFM: information management and stakeholder engagement. The primary issue in information management is the unreliable nature of data sources. Traditional centralized systems frequently struggle to maintain data integrity and accuracy due to a lack of architecture and framework for data transparency. Additionally, they noted that the main challenges in stakeholder engagement stem from a lack of trust and transparency in communication among the parties involved. These challenges and limitations emphasize the need for innovative solutions to enhance data integrity and improve the transparency and efficiency of coordination processes within CbFM to realize more sustainable, efficient, community-centric, and decentralized facility management practices.

### 2.2. Overview of Decentralized Autonomous Organization

A Decentralized Autonomous Organization (DAO) is a novel form of organizational structure that leverages blockchain technology and smart contracts to operate without traditional hierarchical management (Ly and Shojaei 2025). Blockchain is a decentralized, distributed ledger technology that records transactions securely and transparently (Naderi et al. 2024). DAOs are characterized by their decentralized nature, where decision-making power is distributed among members rather than concentrated in a central authority (Rikken et al. 2023). These organizations function autonomously through pre-programmed rules encoded in smart contracts, ensuring transparency as all transactions and governance processes are recorded on the blockchain network.

The core principles of DAOs revolve around three key concepts: decentralization, autonomy, and automation (Ly and Shojaei 2025). Unlike traditional top-down management structures, DAOs utilize a distributed network architecture, eliminating the need for a central authority. Governance in DAOs is achieved through a collaborative process, with community members actively engaging in and voting on various initiatives. Once approved, these initiatives are automatically implemented by smart contracts, ensuring consistent and transparent execution of collective choices. Moreover, the inherent characteristics of blockchain technology enable DAOs to automate organizational procedures and transactions via pre-established protocols encoded in smart contracts, thereby enhancing trust and accountability.

### 2.3. Decentralized autonomous organization in the construction industry

In recent years, multiple studies have highlighted the capabilities of DAOs in enabling decentralized coordination of project management processes. For instance, a study by Spychiger et al. (2023) introduced the concept of a Decentralized Autonomous Project Organization (DAPO), an Ethereum-based platform designed to explore the impact of blockchain and DAO technologies on traditional project management approaches. Additionally, Darabseh and Poças Martins developed a prototype of a decentralized governance system for construction projects using the Aragon platform. Furthermore, Dounas et al. proposed an innovative, collaborative architectural design system called ArchiDAO (Dounas et al. 2022), which combines stigmergic principles, blockchain immutability, and decentralized governance to foster collaboration and shared ownership in design processes. Similarly, Dounas and Lombardi (2019) proposed a reputation system-based DAO for architectural design with shape grammars within a decentralized application (DApp) (Cai et al. 2018).

However, there is a notable knowledge gap in the understanding of decentralized governance in the context of physical infrastructure such as smart building facility management. Ly et al. (2024) proposed a conceptual framework integrating digital twins and decentralized autonomous organizations (DAOs) for smart building facilities management. In their subsequent work, Ly et al. (2024) expanded on this concept by introducing a Decentralized Autonomous Building Cyber-Physical System that integrates DAOs, Large Language Models (LLMs), and digital twins to create a self-managed and financially autonomous building infrastructure. This framework enhances decentralized decision-making and operational adaptability by using LLM-based AI assistants for intuitive human-building interactions. However, the research on DAO governance applications specifically for data and community-driven facility management in smart buildings remains largely unexplored.

### 2.4. Data-driven facilities management

Data-driven facilities management refers to the practice of using data collection, analysis, and insights to optimize the operation, maintenance, and strategic planning of buildings and infrastructure. It leverages various data sources (e.g., sensors, building systems, occupancy patterns) and advanced technologies such as IoT, digital twin, data analytics, and artificial intelligence, to improve building operational efficiency and enhance occupant comfort and satisfaction in the building infrastructure. These technologies support data-driven decision support and automation, allowing facility managers to make more informed decisions about operations, maintenance schedules, space utilization, and energy efficiency initiatives.

The application of digital twin technology in facilities management has shown promising results for decision support and operational optimization. For instance, research by Bujari et al. (2021) introduced the Interactive Planning Platform for City District Adaptive Maintenance Operations (IPPODAMO), a digital twin solution that aims to enhance the Urban Facility Management (UFM) process by quantifying activity levels in areas of interest and considering potential interference from various urban stakeholders. In a related study, Seghezzi et al. (2021) explored the development of an occupancy-oriented digital twin for facility management to enhance building space management. Beyond the building facilities management context, Chen et al. (2024) contributed to the digital twin-driven management research by developing a mixed reality-based digital twin prototype for air traffic management operations.

AI and machine learning significantly enhance facilities management by enabling data-driven decision-making and automation. They address challenges in fault detection, occupancy prediction, energy management, anomaly detection, and maintenance scheduling. For example, Mutis et al.(2020) used a deep neural network with YOLO V3 for occupancy detection, achieving 10–15% energy savings. Predictive maintenance using LSTM models helps detect system anomalies early (Jung et al. 2021), while clustering methods like K-means identify irregular energy usage patterns (Mascali et al. 2023). Cheng et al. (2020) proposed a BIM- and IoT-based framework for MEP maintenance. AI and digital twins, as shown by Zheng et al. (2024), also support real-time tunnel fire safety. However, traditional AI models depend heavily on large, high-quality datasets (Hong et al. 2020; Sanzana et al. 2022). These models typically operate within the bounds of their training data and predefined rules, lacking the flexibility and contextual understanding necessary for more nuanced decision-making.

The emergence of large language models (LLMs) and their ability to understand and generate human-like responses open new opportunities to enhance data-driven management. LLMs could potentially enable more natural language interaction for querying and analyzing digital twin data and provide decision support in complex facility management scenarios. Despite these promising capabilities, there is a notable gap in the current literature as previous studies have yet to explore the integration of LLMs with advanced technology such as digital twins and IoT for data-driven governance in facilities management.

### 2.5. Application of LLM in AEC

Recently, large language models (LLMs) have also attracted significant interest from researchers in the construction domain. This has led to multiple studies exploring their application across various phases of the construction project lifecycle, including planning, construction, operation, and maintenance. For instance, Prieto et al. [27] examined the use of LLMs like ChatGPT for generating construction schedules based on project scopes and requirements. Additionally, You et al. [28] introduced RoboGPT, which leverages ChatGPT's reasoning capabilities for automated sequence planning in robot-based assembly tasks. Furthermore, Chen et al. [29] proposed the Visual Construction Safety Query (VCSQ) system, integrating real-time image captioning and visual question-answering on AR devices using ChatGPT 4. In another study, Uddin et al. [30] investigated the impact of integrating ChatGPT into the construction education curriculum. Also, Zheng and Fischer [31] developed BIMS-GPT, a virtual assistant framework using GPT technologies like ChatGPT for natural language-based searches of building information models (BIMs).

While previous research has explored the integration of GPT models in various construction domains, the application of LLMs in facilities management and facilitating governance tasks specifically in smart building infrastructure remains underexplored. In addition, it's important to note that existing studies in the construction industry predominantly utilize commercialized GPT models like OpenAI's ChatGPT. Researchers have highlighted several concerns associated with these models' usage including data privacy, cost, scalability, and confidentiality (Saka et al. 2024). Therefore, it is necessary to explore alternative options in inferencing these generative models, which include the use of local and open-source large language models. These alternatives can enhance processing speed and address privacy issues by keeping data within the device or system.

## 3. Research Methodology

This study followed the Design Science Research approach (Geerts 2011), a rigorous methodological framework that creates innovative artifacts, such as models, algorithms, or frameworks, to solve practical real-world problems, evaluate the designed solutions, and contribute to the existing knowledge base through a structured and iterative process. Fig. 1 illustrates the DSR-based research methodology used in this study. The DSR methodology has been used by many researchers in the construction industry for the development of blockchain-related frameworks and applications (Erri Pradeep et al. 2021; Tao et al. 2023). There are 6 steps in the DSR methodology:

(1) Problem identification and motivation. The literature review in section 2 reveals several research gaps in the data-driven management of building infrastructure. These include the lack of decentralized frameworks for transparent and inclusive decision-making in facilities management, limited studies on the integration of LLM and digital twin, lack of the LLM-assisted governance framework for building FM, and the need for the open-source LLM application in the construction industry especially in the smart building domain.

(2) Objective definition. To fulfill these knowledge gaps, this study aims to design and develop an AI and data-driven distributed governance framework for building facilities management by integrating large language models, digital twins, and decentralized autonomous organizations.

(3) Design and development. The design and development of the proposed framework will be achieved through the following objectives (i) To develop a DAO-based decentralized FM governance platform (ii) To create a blockchain-based decentralized framework for building facilities operation (iii) To develop an LLM-based AI agents for FM governance assistance through the integration with digital twin data. (iv) To evaluate the system performance in real-world case studies.

(4) Demonstration. The developed framework and prototypes will be deployed on the actual physical building. Different functionalities and use cases of the system will be demonstrated including the use cases for a decentralized governance platform for FM, AI assistant, and digital twin usage in assisting governance decisions.

(5) Evaluation. Both quantitative measures and qualitative assessments of the framework will be provided.

(6) Communication. A prototype of the system will be demonstrated to relevant stakeholders, including facility managers, building owners, and occupants to receive feedback and further improvement. The methodology for the development of the proposed system and evaluation results will be published in international academic journals.

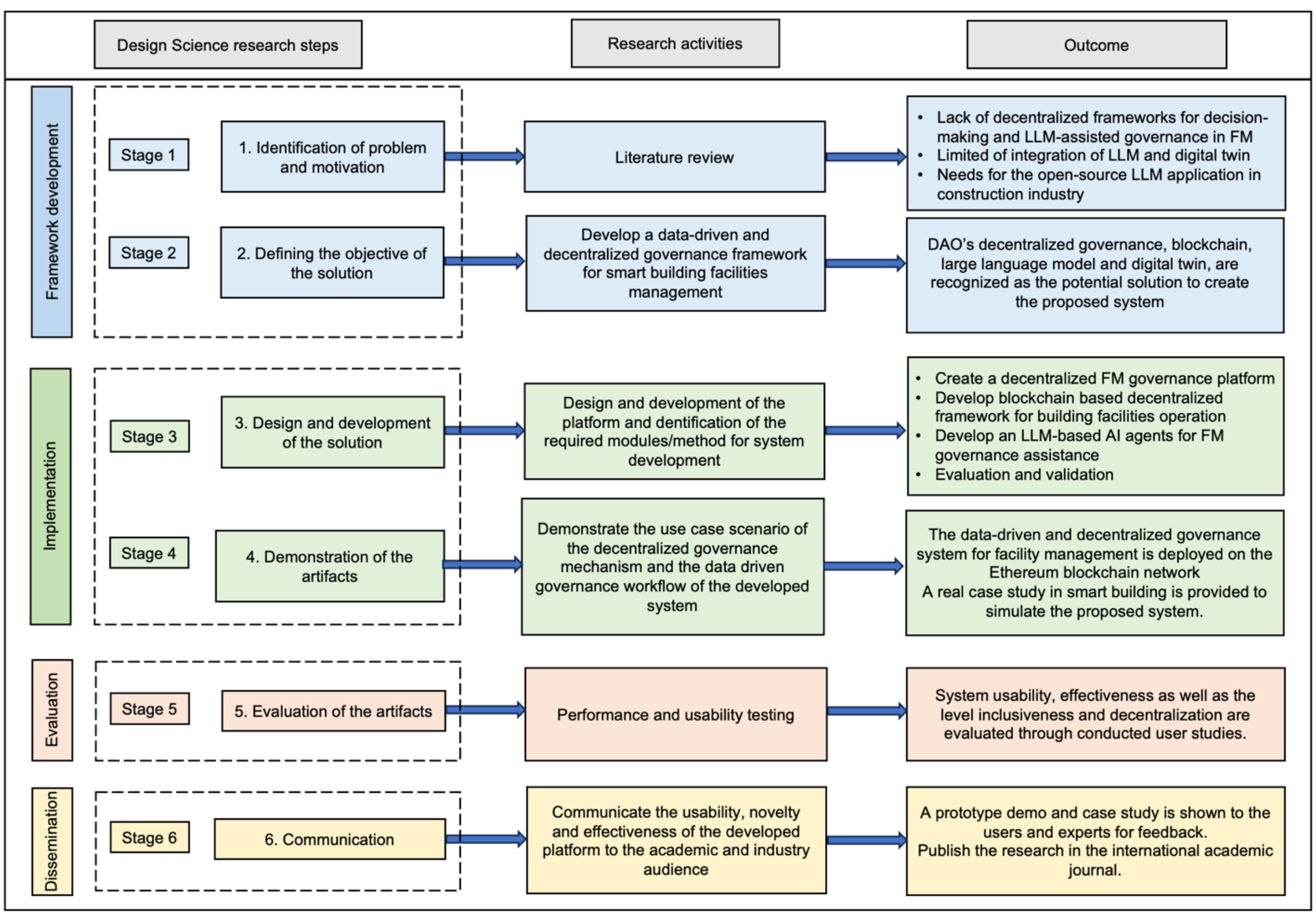

**Fig. 1** Design Science Research-driven research flow

## 4. Proposed framework

### 4.1. Framework overview

The objective of this study is to create a data-driven and decentralized governance platform that empowers stakeholders to collaboratively participate in the management and coordination of building operations transparently and democratically through blockchain-based DAO, digital building twin, and large language model-based AI assistant. The architecture of the proposed framework and the relationship between its main components are illustrated in Fig. 2. The framework is mainly comprised of the physical component and cyber components. At the core of the physical component are the building assets equipped with IoT devices and sensors. These sensors continuously monitor environmental conditions such as temperature, humidity, light, gas, and motion. A Raspberry Pi serves as the interface between the physical sensors and the cyber system, processing and transmitting data to the cyber world. The key elements within the cyber component include decentralized storage, decentralized application (DApp), blockchain network, and digital twin. The cyber component of the framework is centered around the DApp which incorporates two other key elements: a DAO-based decentralized governance platform and an LLM-based AI assistant platform. The governance platform enables stakeholders or DAO members to propose, vote on, and implement facility management policies through a transparent and democratic process.

The first element of the data-driven decision support workflow is the digital twin, which provides real-time visualization of the physical space. The second element of the decision support is the LLM-based AI assistant which allows DAO members to query and analyze data from the historical records. It generates insights and suggestions to aid DAO members in making informed facility management decisions. This AI-driven approach allows for more sophisticated analysis of complex building performance metrics, energy consumption patterns, and occupancy data.

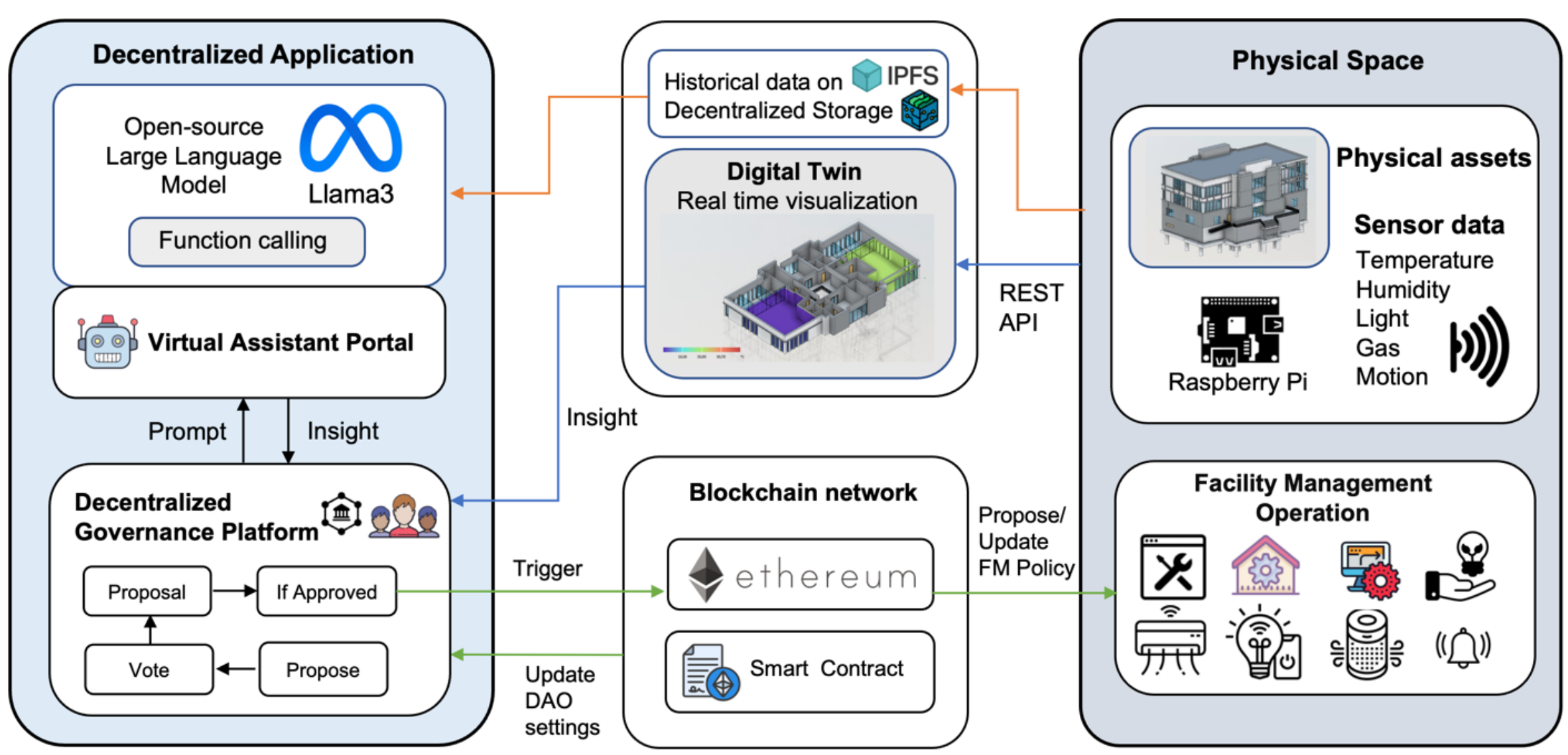

**Fig. 2** Overview of the proposed framework

### 4.2. Decentralized governance platform

Decentralized governance platform is the primary component of the proposed framework, which is responsible for the decision-making on the rules and policy of the building facilities management process (Fig. 3). Depending on the functions of the building, the governance platform in this study will be specifically customized/designed for different groups of key and high-level stakeholders in the building including building owners, building facility managers, administrative personnel, faculty, etc. These decisions by DAO members are informed by real-time data visualizations through the live digital twin of the building infrastructure and analysis of historical data stored on decentralized storage made by the LLM-based AI assistant. This governing body can leverage the transparency and collective decision-making capabilities of a DAO to democratize the governance of core building operations including automation of building operations, and settings of HVAC or lighting systems. This data-driven approach allows stakeholders to make informed decisions based on actual building performance metrics, energy consumption patterns, and occupancy data. This design can decentralize the control of these high-stakes decisions from a single centralized authority while also offering oversight from authorized entities with comprehensive knowledge and accountability.

Difference governance-related parameters such as eligibilities of the voters, the available voting options, the voting period, voting delay, and the quorum or threshold requirements and tokens distributions and voting settings will be set before the initial deployment of the DAO. Upon its initial deployment, the predetermined amount of governance tokens will be minted to the corresponding stakeholders or DAO members. The minted governance tokens play a pivotal role in determining voting rights and influence within the decision-making processes. Greater allocation of tokens will grant DAO members more voting power and governance rights compared to their peer members. In addition, the study adopts the ERC-20 token standard (Cuffe 2018) for governance tokens. In addition, the developed DAO platform employs a token-based quorum voting mechanism, where the weight of voting power corresponds to the number of tokens held (Fan et al. 2023).

DAO governance mechanism consists of several stages from proposal submission and voting to queue and execution. It begins with the proposal stage, where a member submits a proposal via the DAO platform. This proposal is encoded with specific details and objectives, typically governed by predefined conditions in the DAO's smart contract. Once submitted, there is often a voting period initiated. During this period, DAO members eligible to vote can review the proposal, engage in discussions, and make informed decisions on whether to support or oppose it. Following the voting period, a voting delay period may be activated. This delay allows time for further deliberation and potentially allows members to retract their votes or challenge the proposal before it becomes active. After the delay period, if the proposal garners sufficient support with the predefined quorum or threshold reached the proposal will be moved into the queue and execution phase. The queue and execution involve implementing the proposal's actions as specified in the smart contract. These actions could include transferring funds, updating new facilities management policy, etc. The execution phase typically includes mechanisms to ensure transparency and accountability, with all actions recorded on the blockchain ledger. These processes ensure that DAO governance operates efficiently, with clear stages from proposal submission through to actionable outcomes based on community consensus.

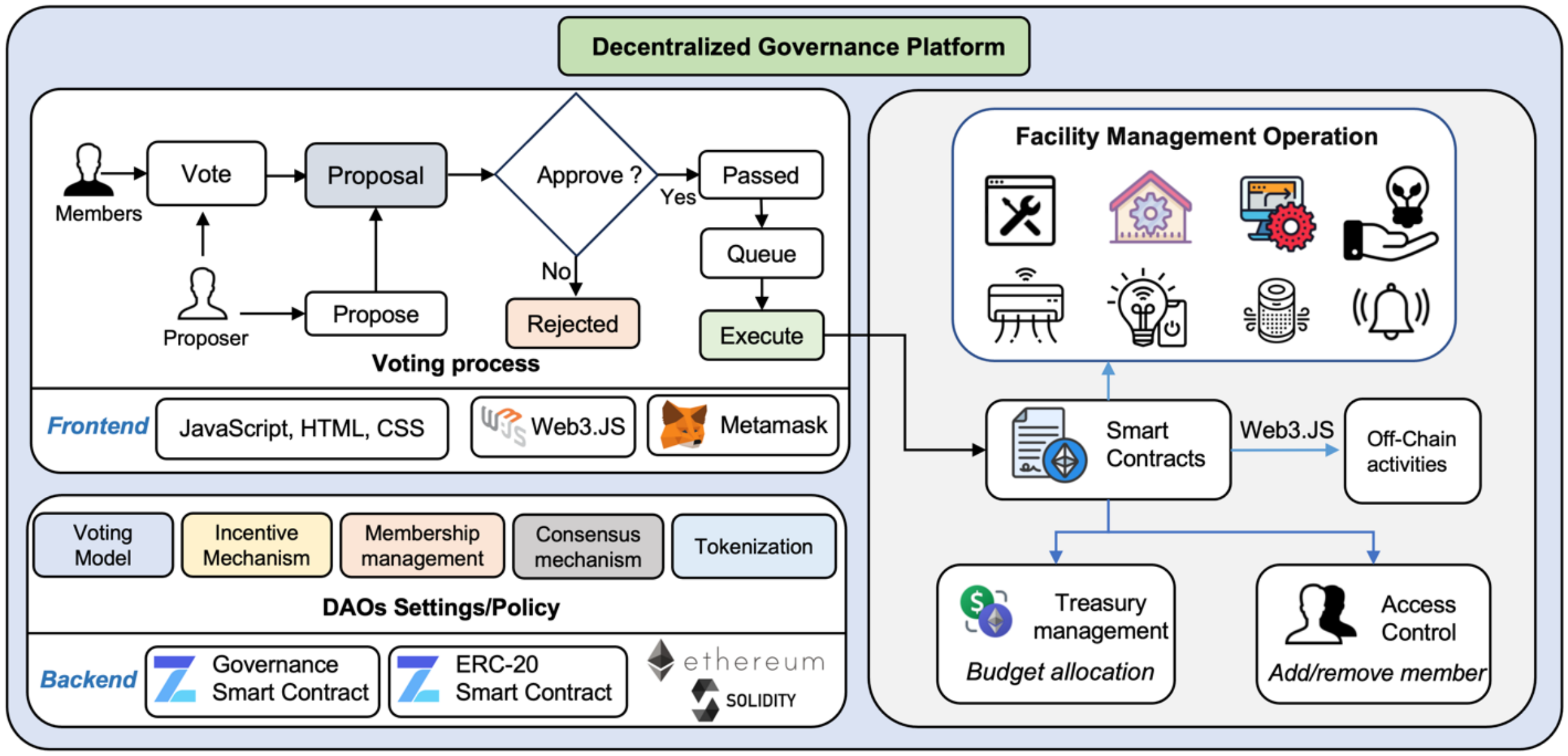

**Fig. 3** Overview of the decentralized governance platform

### 4.3. Digital building twin

Digital building twin is one of the core components of the data-driven approach for facilities management by offering a comprehensive virtual representation of the physical structure. This digital twin setup enables building managers or DAO members from the decentralized governance platform to gain instant insights into the building's environmental conditions, energy usage, and occupancy patterns. This comprehensive, real-time view of the building's status will inform the building administrator or DAO member of the building conditions as well as the analytics of the building usage which facilitates the data-driven decisions building facilities management strategies, allowing them to make informed decisions on related task such as energy conservation, space utilization and more. This system integrates various data sources to create a real-time visualization of environmental conditions within a building. The digital building twin development in this study will be provided in section 5.1.2.

### 4.4. AI Assistant for Data Analysis

This section details the usage of an AI Virtual Assistant for Conversational Data Analysis within the framework of data-driven governance for smart buildings using LLM and DAO. This AI assistant serves as a crucial interface between the decentralized governance platform and the sensor data collected from the building environment by offering data analytics and visualization capabilities. As illustrated in Fig. 4, the architecture of this AI assistant is comprised of both frontend and backend components. This interface allows DAO members to interact with the AI assistant through natural language prompts with both textual and voice input, facilitating intuitive data querying and analysis requests.

The AI-assisted data analysis workflow begins once the input prompt and IoT data are submitted by the DAO member to the chat interface. This query is then directed to the LLM backend. Upon receiving a query, the AI assistant employs its reasoning capabilities to interpret the request and execute appropriate data analysis tools. These tools, as shown in the figure, include functions for plotting graphs of environmental or occupancy data, summarizing and analyzing time series data such as temperature, humidity, luminosity, gas levels, occupancy, and energy consumption metrics, before providing suggestions on facility management policies. The results of these analyses are then visualized and presented back to the DAO member through the chat interface. DAO members can also continuously ask the AI agents on the visualization to provide comparisons or insights to facilitate management decisions.

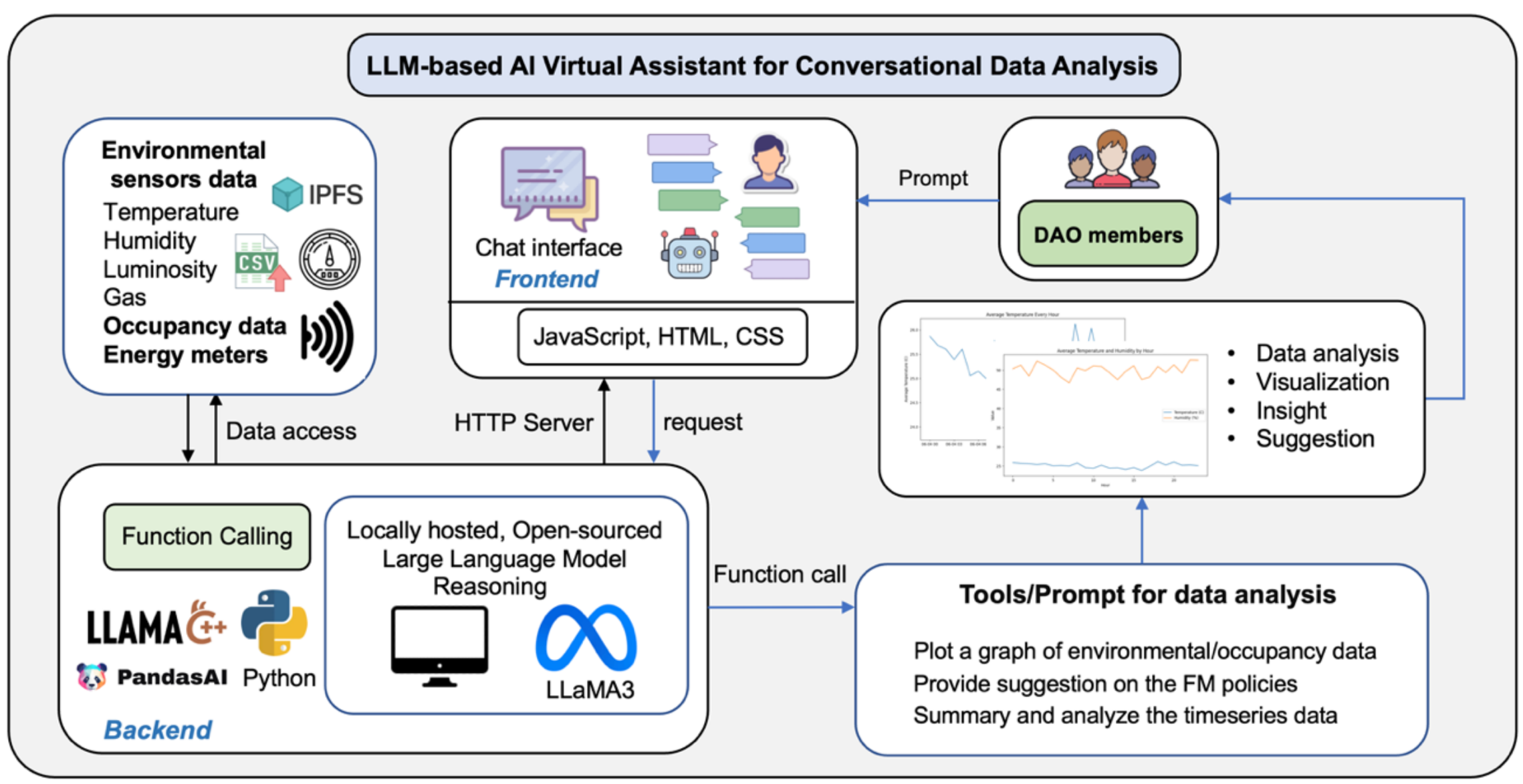

**Fig. 4** Workflow of the AI Virtual Assistant

### 4.5. Data-driven governance

As illustrated in Fig. 5, the data-driven governance workflow in this framework is comprised of AI-powered data analytics with digital twin technologies. This synergy between the AI assistant and the digital twin creates a comprehensive decision support system that can optimize facility management and operational efficiency. The digital building twin provides real-time visualization of environmental data such as temperature, humidity, light intensity, occupancy levels, and CO concentration, offering an intuitive, three-dimensional representation of the building's current state. Simultaneously, the AI assistant conducts an in-depth analysis of these historical and current data, generating graphs and visualization, insights, and suggestions.

One primary use case LLM for data-driven governance is occupancy-based control. For instance, in a scenario where the AI assistant identifies consistently low occupancy in a certain location during a certain hour, despite high energy consumption from HVAC systems and other appliances, it can present this information through clear visualizations and data analysis to the DAO member. The digital building twin can be used in conjunction with the AI system to provide real-time occupancy levels and other environmental conditions. DAO members then can propose and vote on scheduling automated adjustments to the setpoints of these appliances or on/off schedules to optimize energy usage without compromising comfort during occupied periods.

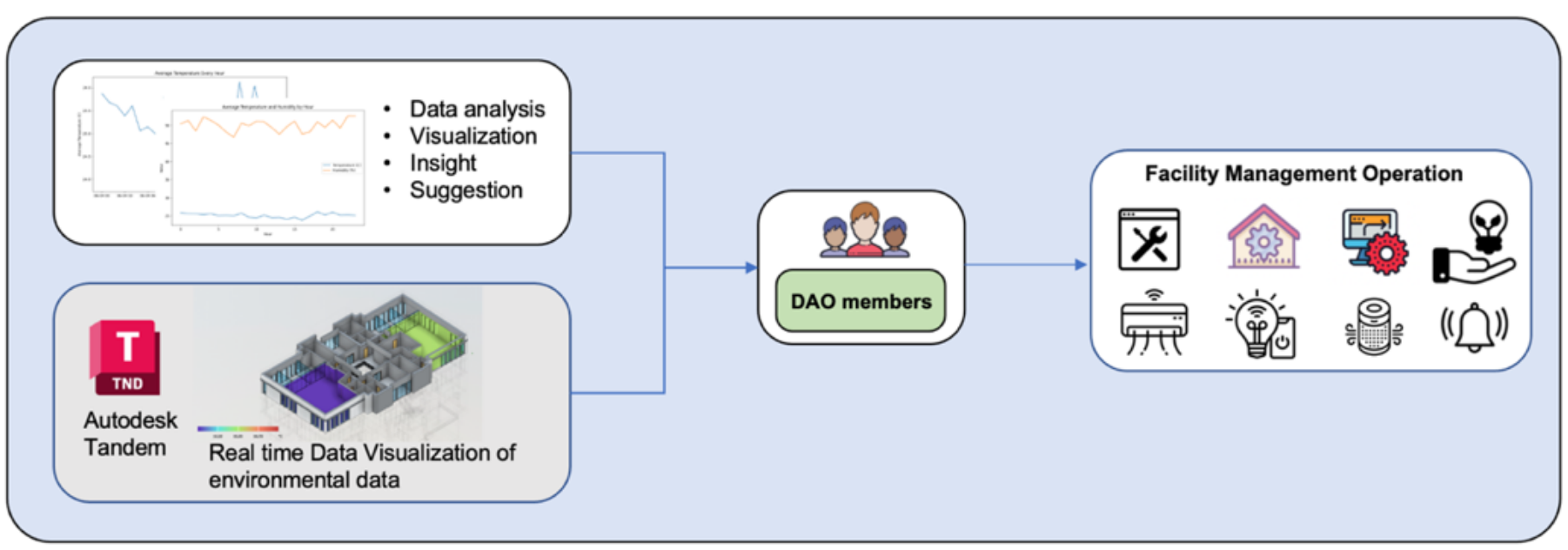

**Fig. 5** Proposed AI and Digital twin-driven decentralized governance workflow

### 4.6. Building operation automation

The proposed framework aims to enable decentralized and automated control of building systems by leveraging the integration of IoT sensors, smart building facilities such as the HVAC and lighting system, and blockchain-based smart contracts. A key feature of this automation framework is the use of blockchain smart contracts to store and manage threshold parameters for automated building operations. These thresholds, which include maximum and minimum values for various environmental factors (e.g. temperature, humidity, alerts, etc.), are established by DAO members who serve as administrators for the building. The use of blockchain technology ensures these parameters are stored securely and transparently, providing an immutable record that forms the foundation of the building's automation logic. During operation, a Python script will be written to continuously compare real-time environmental sensor data and building operation metrics as described in the previous section against these predefined, thresholds from the blockchain's smart contract. When environmental values exceed the predefined limits, the system triggers specific actions in the building's smart systems, such as adjusting HVAC setpoints, modifying lighting conditions, activating, or deactivating air quality management devices, and triggering alerts to maintain optimal building conditions.

## 5. Proof of concept

In this section, a case study with the developed prototypes is used to validate the viability and functionality of the framework. The code for the technical implementation of the prototypes is available under an open-source license (Ly 2024, 2025). The tools, coding languages, and development environments employed for each module of the prototypes are summarized in Table 1.

**Table 1** Tools used for prototype development

| Tasks | Programming language (packages) | Development environment |
|---|---|---|
| Frontend web pages development | React JS | Visual Studio Code |
| Smart contract development | Solidity | Brownie |
| Digital building twin | JavaScript (Autodesk API) | Visual Studio Code |
| IoT sensors and smart home device | Python | Visual Studio Code |
| Interaction between Dapp and smart contract | JavaScript (web3.js API) | Visual Studio Code |
| Large language models deployment | C++ (llamacpp) | Visual Studio Code |
| Large language models Inference | Python (llamacpp-python) | Visual Studio Code |

### 5.1 Development of the Dapp backend

#### 5.1.1. Smart contract design and development

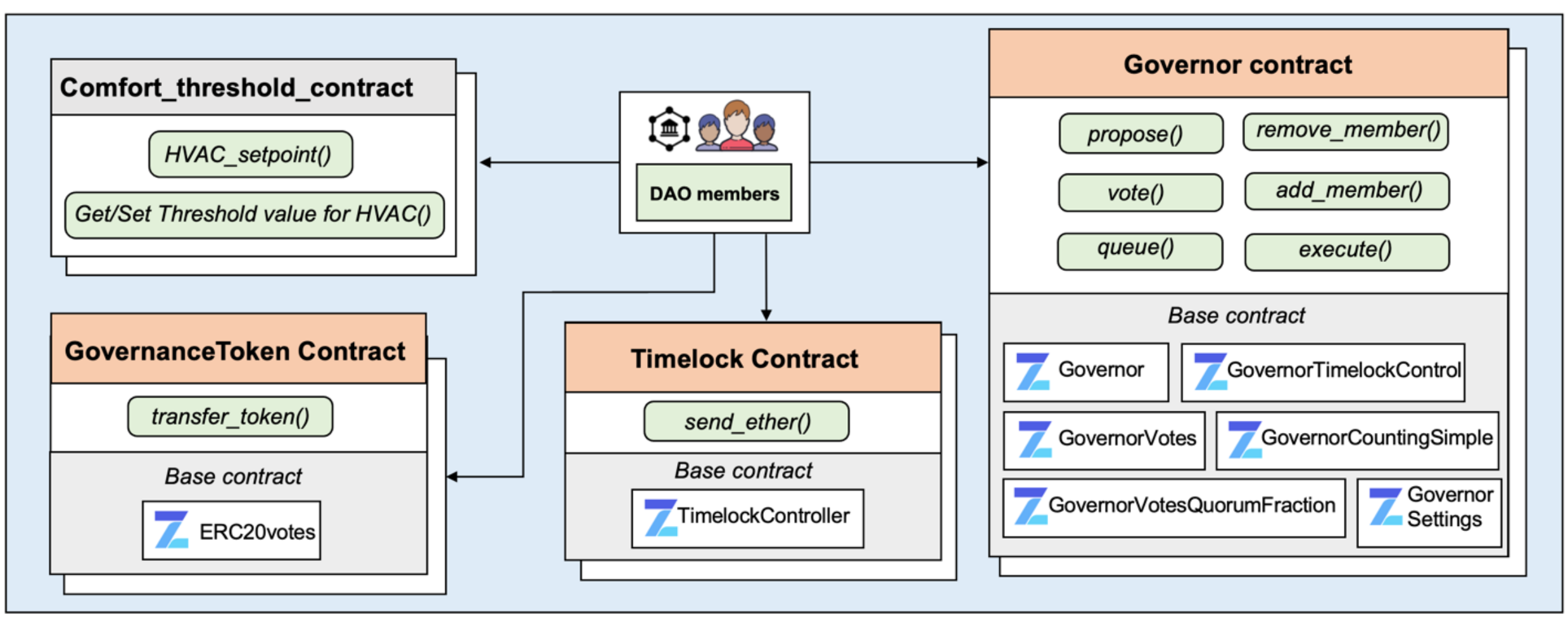

**Fig. 6** Design of the smart contracts and their relationship between actors in the systems

The proposed DAO-based data-driven governance framework incorporates a set of smart contracts that form the backbone of its decentralized application components. Fig. 6 illustrates the high-level design and interrelationships of these contracts, which include the decentralized governance platform and environmental comfort threshold contract for building facilities operations. Central to the decentralized governance platform lie three core smart contracts: the Governor contract, the Timelock contract, and the GovernanceToken contract. These contracts are built upon the robust and widely adopted OpenZepplin library which is renowned for its security and adherence to community standards in DAO development. The DAO governor contract incorporates multiple OpenZeppelin base contracts to deliver a robust governance infrastructure (Fig. 7). Specifically, the Governor base contract enables proposal generation and execution, while GovernorTimelockControl enhances security by introducing an execution delay for approved proposals. Voting capabilities are linked to ERC-20 governance tokens through the GovernorVotes component. Additionally, GovernorCountingSimple provides the methodologies for vote calculation, and GovernorVotesQuorumFraction enforces participation thresholds relative to the total token supply. The GovernorSettings component allows for the configuration of key governance parameters such as voting timeframes and waiting periods. The DAO Governor contract operates in conjunction with the Timelock contract to program execution sequences for approved proposals and interacts with the Governance Tokens contract to authenticate member voting authority during the proposal process.

The Governance Tokens contract manages fungible tokens representing voting power within the framework (Fig.8 a). Built upon the ERC20Votes base contract, it provides token-based voting and delegation capabilities using standard ERC20 tokens. Token holders can vote directly or delegate their voting power to other members using the transfer_token function. The Timelock contract introduces a mandatory delay between proposal approval and execution (Fig.8 b). Extending from the TimelockController base contract, it queues approved proposals for a specified period, allowing stakeholders time for review. This contract manages the DAO's monetary assets and includes functions like send_ether for transactions between the DAO and external addresses. The framework also includes a Comfort_Threshold_contract which contains functions for managing building operations through set and get functions for threshold variables (Fig.8 c). These allow DAO members to define parameters for smart building facilities operations, accessible via Python scripts using the web3 JS library.

```
11   contract DABCPSGovernor is
12       Governor,
13       GovernorSettings,
14       GovernorCountingSimple,
15       GovernorVotes,
16       GovernorVotesQuorumFraction,
17       GovernorTimelockControl
18   {
19       // Proposal Counts
20       uint256 public s_proposalCount;
21       address[] public members;
22       mapping(address => bool) public isMember;
23       constructor(
24           IVotes _token,
25           TimelockController _timelock,
26           uint256 _votingDelay,
27           uint256 _votingPeriod,
28           uint256 _quorumPercentage
29       )
30           Governor("Governor")
31           GovernorSettings(
32               _votingDelay, /* 1 => 1 block */
33               _votingPeriod, /* 300 blocks => 1 hour */
34               0 /* 0 => Because we want anyone to be able to create a proposal */
35           )
36           GovernorVotes(_token)
37           GovernorVotesQuorumFraction(_quorumPercentage) /* 4 => 4% */
38           GovernorTimelockControl(_timelock)
39       {
40           s_proposalCount = 0;
41           members.push(msg.sender);
42           isMember[msg.sender] = true;
43       }
44       fallback() external payable {}
45       modifier onlyMember() {
46           require(isMember[msg.sender], "Only members can call this function");
47           _;
48       }
49       function sendEther(address payable receiver, uint256 amount) external {
50           require(address(this).balance >= amount, "Insufficient balance in the contract");
51           receiver.transfer(amount);
52       }
53       function getMembers() external view returns (address[] memory) {
54           return members;
55       }
56       // The following functions are overrides required by Solidity.
57       function addMember(address newMember) external onlyMember {
58           require(!isMember[newMember], "Address is already a member");
59           members.push(newMember);
60           isMember[newMember] = true;
61       }
62       function removeMember(address member) external onlyMember {
63           require(isMember[member], "Address is not a member");
64
65           // Find the index of the member in the array
66           uint256 index;
67           for (uint256 i = 0; i < members.length; i++) {
68               if (members[i] == member) {
69                   index = i;
70                   break;
71               }
72           }
73           // Swap with the last element and then remove the last element to maintain order
74           members[index] = members[members.length - 1];
75           members.pop();
76           isMember[member] = false;
77       }
78       function getMemberLength() external view returns (uint256) {
79           return members.length;
80       }
81       function votingDelay()
82           public
83           view
84           override(IGovernor, GovernorSettings)
85           returns (uint256)
86       {
87           return super.votingDelay();
88       }
```

```
97       function votingPeriod()
98           public
99           view
100          override(IGovernor, GovernorSettings)
101          returns (uint256)
102      {
103          return super.votingPeriod();
104      }
105      function quorum(uint256 blockNumber)
106          public
107          view
108          override(IGovernor, GovernorVotesQuorumFraction)
109          returns (uint256)
110      {
111          return super.quorum(blockNumber);
112      }
113      function state(uint256 proposalId)
114          public
115          view
116          override(Governor, GovernorTimelockControl)
117          returns (ProposalState)
118      {
119          return super.state(proposalId);
120      }
121      function propose(
122          address[] memory targets,
123          uint256[] memory values,
124          bytes[] memory calldatas,
125          string memory description
126      ) public override(Governor, IGovernor) returns (uint256) {
127          s_proposalCount++;
128          return super.propose(targets, values, calldatas, description);
129      }
130      function proposalThreshold()
131          public
132          view
133          override(Governor, GovernorSettings)
134          returns (uint256)
135      {
136          return super.proposalThreshold();
137      }
138      function _execute(
139          uint256 proposalId,
140          address[] memory targets,
141          uint256[] memory values,
142          bytes[] memory calldatas,
143          bytes32 descriptionHash
144      ) internal override(Governor, GovernorTimelockControl) {
145          super._execute(proposalId, targets, values, calldatas, descriptionHash);
146      }
147      function _cancel(
148          address[] memory targets,
149          uint256[] memory values,
150          bytes[] memory calldatas,
151          bytes32 descriptionHash
152      ) internal override(Governor, GovernorTimelockControl) returns (uint256) {
153          return super._cancel(targets, values, calldatas, descriptionHash);
154      }
155      function _executor()
156          internal
157          view
158          override(Governor, GovernorTimelockControl)
159          returns (address)
160      {
161          return super._executor();
162      }
163      function supportsInterface(bytes4 interfaceId)
164          public
165          view
166          override(Governor, GovernorTimelockControl)
167          returns (bool)
168      {
169          return super.supportsInterface(interfaceId);
170      }
171      function getNumberOfProposals() public view returns (uint256) {
172          return s_proposalCount;
173      }
174  }
```

**Fig. 7** Decentralized Governance Platform’s DAO Governor contract

```
5   pragma solidity ^0.8.7;
6
7   contract GovernanceToken is ERC20Votes {
8       // events for the governance token
9       event TokenTransfered(
10          address indexed from,
11          address indexed to,
12          uint256 amount
13      );
14      // Events
15      event TokenMinted(address indexed to, uint256 amount);
16      event TokenBurned(address indexed from, uint256 amount);
17      // max tokens per user
18      uint256 constant TOKENS_PER_USER = 2000;
19      uint256 constant TOTAL_SUPPLY = 1000000 * 10**18;
20      uint256 public data;
21      // Mappings
22      mapping(address => bool) public s_claimedTokens;
23      // Number of holders
24      address[] public s_holders;
25      constructor(uint256 _keepPercentage)
26          ERC20("BFHTOKEN", "BFHT")
27          ERC20Permit("BFHToken")
28      {
29          uint256 keepAmount = (TOTAL_SUPPLY * _keepPercentage) / 100;
30          _mint(msg.sender, TOTAL_SUPPLY);
31          _transfer(msg.sender, address(this), TOTAL_SUPPLY - keepAmount);
32          s_holders.push(msg.sender);
33      }
34      function sendTokens(address payable receiver, uint256 amount) external  {
35          _transfer(address(this), receiver, amount * 10**18);
36      }
37      function reward(uint256 amount) external  {
38          _transfer(address(this), msg.sender, amount * 10**18);
39      }
40      function getHolderLength() external view returns (uint256) {
41          return s_holders.length;
42      }
43      // Overrides required for Solidiy
44      function _afterTokenTransfer(
45          address from,
46          address to,
47          uint256 amount
48      ) internal override(ERC20Votes) {
49          super._afterTokenTransfer(from, to, amount);
50          emit TokenTransfered(from, to, amount);
51      }
52      function _mint(address to, uint256 amount) internal override(ERC20Votes) {
53          super._mint(to, amount);
54          emit TokenMinted(to, amount);
55      }
56      function _burn(address account, uint256 amount)
57          internal
58          override(ERC20Votes)
59      {
60          super._burn(account, amount);
61          emit TokenBurned(account, amount);
62      }
63  }
```

a)

```
1   //SPDX-License-Identifier: MIT
2
3   pragma solidity ^0.8.9;
4
5   import "@openzeppelin/contracts/governance/TimelockController.sol";
6
7   contract TimeLock is TimelockController {
8       constructor(
9           uint256 minDelay,
10          address[] memory proposers,
11          address[] memory executors,
12          address admin
13      ) TimelockController(minDelay, proposers, executors, admin) {}
14
15      function sendEther(address payable receiver, uint256 amount) external {
16          require(address(this).balance >= amount, "Insufficient balance in the contract");
17          receiver.transfer(amount);
18      }
19
20  }
```

b)

```
1   // SPDX-License-Identifier: MIT
2   pragma solidity ^0.8.0;
3
4   contract SmartBuildingAutomation {
5
6       // Variables for controlling the smart building facilities
7       int256 public minTemperature;
8       int256 public maxTemperature;
9       uint256 public minCO2Level;
10      uint256 public maxCO2Level;
11      uint256 public minLuxLevel;
12      uint256 public maxLuxLevel;
13      uint256 public minHumidity;
14      uint256 public maxHumidity;
15
16      address public constant dao = 0x3aF5647E366fb51C89e4c43Bc8C173dAa018AFf6;
17
18      // Events to notify when values are updated
19      event MinTemperatureUpdated(int256 minTemperature);
20      event MaxTemperatureUpdated(int256 maxTemperature);
21      event MinCO2LevelUpdated(uint256 minCO2Level);
22      event MaxCO2LevelUpdated(uint256 maxCO2Level);
23      event MinLuxLevelUpdated(uint256 minLuxLevel);
24      event MaxLuxLevelUpdated(uint256 maxLuxLevel);
25      event MinHumidityUpdated(uint256 minHumidity);
26      event MaxHumidityUpdated(uint256 maxHumidity);
27
28      // Modifier to restrict access to DAO
29      modifier onlyDAO() {
30          require(msg.sender == dao, "Only DAO can set the values");
31          _;
32      }
33      // Setter functions
34      function setMinTemperature(int256 _minTemperature) public onlyDAO {
35          minTemperature = _minTemperature;
36          emit MinTemperatureUpdated(_minTemperature);
37      }
38      function setMaxTemperature(int256 _maxTemperature) public onlyDAO {
39          maxTemperature = _maxTemperature;
40          emit MaxTemperatureUpdated(_maxTemperature);
41      }
42      function setMinCO2Level(uint256 _minCO2Level) public onlyDAO {
43          minCO2Level = _minCO2Level;
44          emit MinCO2LevelUpdated(_minCO2Level);
45      }
46      function setMaxCO2Level(uint256 _maxCO2Level) public onlyDAO {
47          maxCO2Level = _maxCO2Level;
48          emit MaxCO2LevelUpdated(_maxCO2Level);
49      }
50      function setMinLuxLevel(uint256 _minLuxLevel) public onlyDAO {
51          minLuxLevel = _minLuxLevel;
52          emit MinLuxLevelUpdated(_minLuxLevel);
53      }
54      function setMaxLuxLevel(uint256 _maxLuxLevel) public onlyDAO {
55          maxLuxLevel = _maxLuxLevel;
56          emit MaxLuxLevelUpdated(_maxLuxLevel);
57      }
58      function setMinHumidity(uint256 _minHumidity) public onlyDAO {
59          minHumidity = _minHumidity;
60          emit MinHumidityUpdated(_minHumidity);
61      }
62      function setMaxHumidity(uint256 _maxHumidity) public onlyDAO {
63          maxHumidity = _maxHumidity;
64          emit MaxHumidityUpdated(_maxHumidity);
65      }
66      // Getter functions
67      function getMinTemperature() public view returns (int256) {
68          return minTemperature;
69      }
70      function getMaxTemperature() public view returns (int256) {
71          return maxTemperature;
72      }
73      function getMinCO2Level() public view returns (uint256) {
74          return minCO2Level;
75      }
76      function getMaxCO2Level() public view returns (uint256) {
77          return maxCO2Level;
78      }
79      function getMinLuxLevel() public view returns (uint256) {
80          return minLuxLevel;
81      }
82      function getMaxLuxLevel() public view returns (uint256) {
83          return maxLuxLevel;
84      }
85      function getMinHumidity() public view returns (uint256) {
86          return minHumidity;
87      }
88      function getMaxHumidity() public view returns (uint256) {
89          return maxHumidity;
90      }
91  }
```

c)

**Fig. 8** Governance tokens and building comfort threshold-related smart contracts (a) Governance token contract (b) Timelock controller contract (c) Comfort threshold contract

### 5.1.2. Digital building twin development

The digital building twin in the framework aims to provide visualization of environmental and occupancy conditions, as well as energy usage. This study uses Bishop-Favrao Hall at Virginia Tech as a case study due to its dynamic occupancy patterns, which are ideal for testing the framework's ability to capture and analyze fluctuating building utilization and environmental conditions. The BIM model of Bishop-Favrao Hall was developed using Autodesk Revit 2024 (Fig. 9). The first digital twin collects environmental data via a Raspberry Pi 4B interfacing with a series of environmental sensors and IoT devices such as DHT11 (Temperature & Humidity) Light Intensity Sensor MQ-2 Gas Sensor Groove Smart Plug (Energy Metering). Sensor data is processed using Python libraries (e.g., Adafruit_DHT, Rpi.GPIO, Adafruit_MQTT) and transmitted to the digital twin platform via a REST API (Flask library). Data is sent in JSON format and integrated into the BIM model using Autodesk Platform Services' Model Derivative API, enabling real-time visualization for DAO members. Fig. 10 outlines the digital twin development workflow.

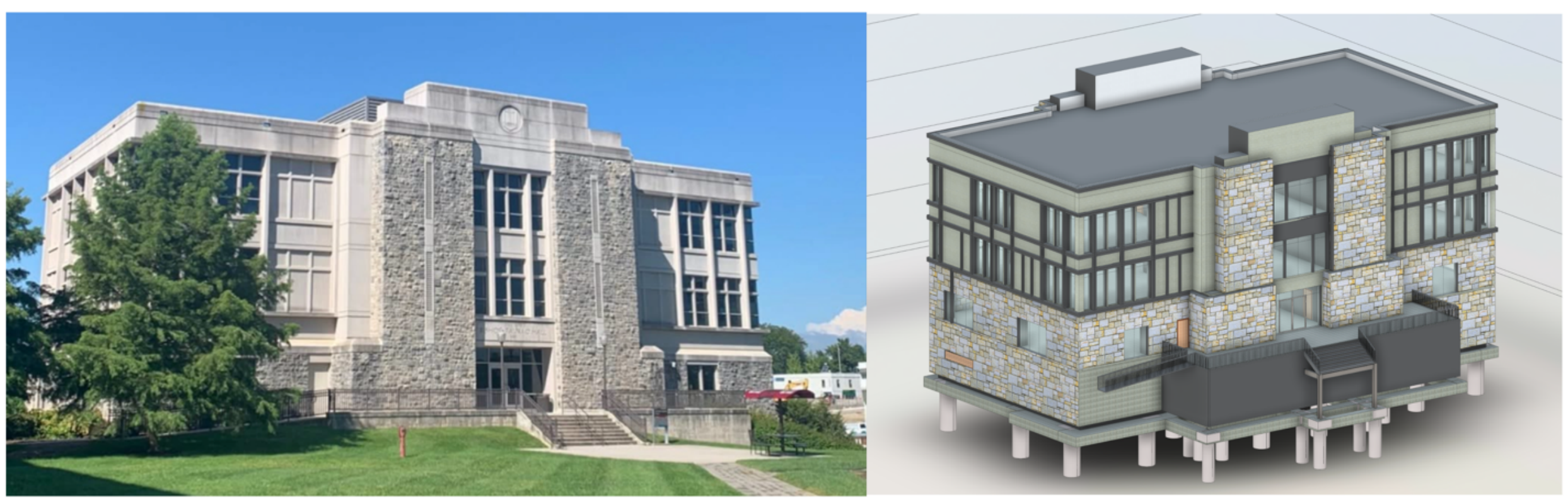

**Fig. 9** Myers-Lawson School of Construction's Bishop-Favrao Hall and its BIM model

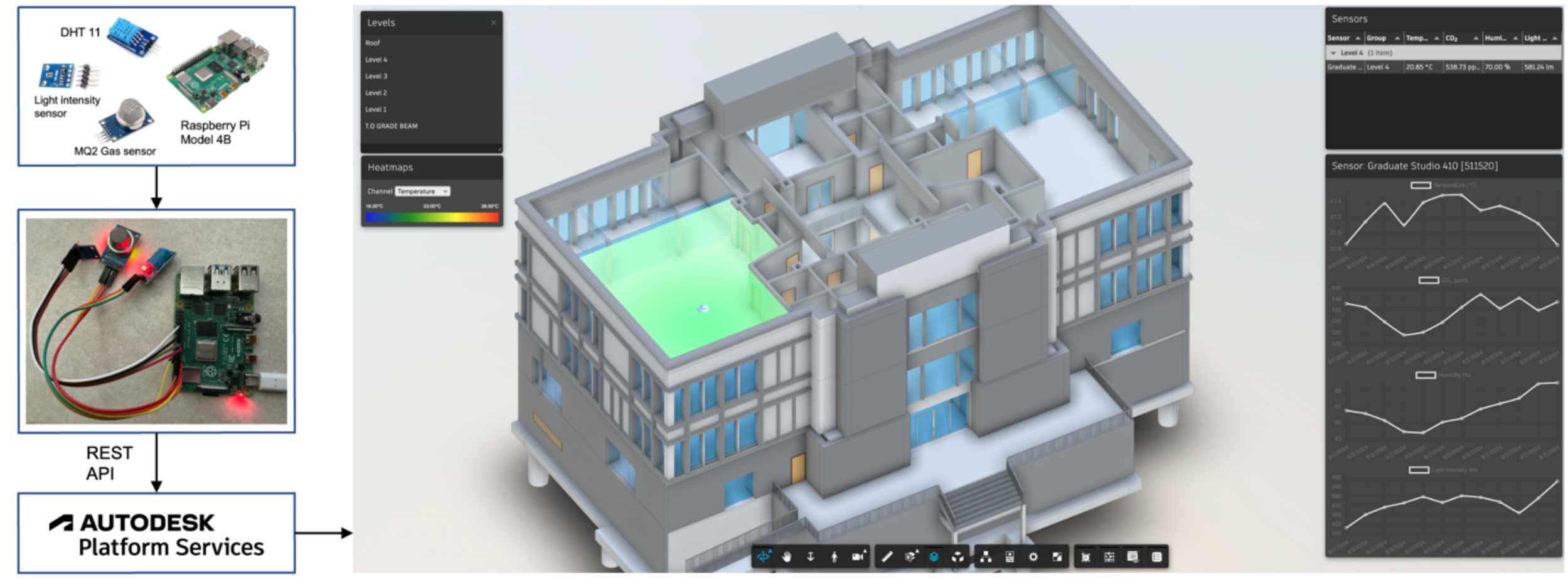

**Fig. 10** Workflow of digital building twin development

### 5.1.3. AI Virtual Assistant development

This study implemented the AI functionality through a virtual assistant, which is powered by Meta's open-source large language model, LLaMA 3 8b (Dubey et al. 2024). The workstation used for the LLaMA 3 model deployment in this study is an Apple MacBook Pro with an M1 Max chip with 32GB of RAM. To run the LLaMA 3 model efficiently, this study employs a quantized version of the model using the llama.cpp library (Gerganov 2024). Llama.cpp is a tool that allows the execution of quantized LLMs on local hardware with support for different types of GPU. Quantization of LLM is a technique to reduce the model's size and computational requirements, aiming to improve inference speed while maintaining lower memory usage (Zhao et al. 2024). This feature is particularly important for deploying the AI assistant on local hardware. The backend also incorporates function calling capabilities, utilizing Python and libraries like PandasAI (2024) for data manipulation and analysis.

### 5.2. Development of the Dapp Frontend

The front end of the Dapp for the proposed system was developed using React JS due to its flexibility, modular structure, and compatibility with web3 JS, which facilitates the interaction between the Ethereum blockchain and the web application. MetaMask was integrated to connect users' Ethereum wallets, enabling blockchain-related transactions. As depicted in Fig. 11, the Dapp interface comprises six primary navigation tabs: Governance, Treasury, Digital Twin, User, Member, and AI Assistant. The Governance section enables DAO participants to submit and vote on operational proposals regarding building systems such as facility automation configurations and asset management. The Treasury tab presents financial information including the DAO's governance token holdings and Ethereum cryptocurrency balance, along with mechanisms to propose and vote on financial transfers. The Digital Twin section offers dual visualization capabilities—one displaying real-time building metrics (occupancy levels, energy consumption, and environmental parameters), while the second presents historical data of these metrics. The AI Assistant tab provides an interface where users can engage with the virtual assistant functionality. The Member tab offers the overview of all individual DAO member and their governance token in the system while the User tab provides the status of the available governance tokens and cryptocurrency of the current user.

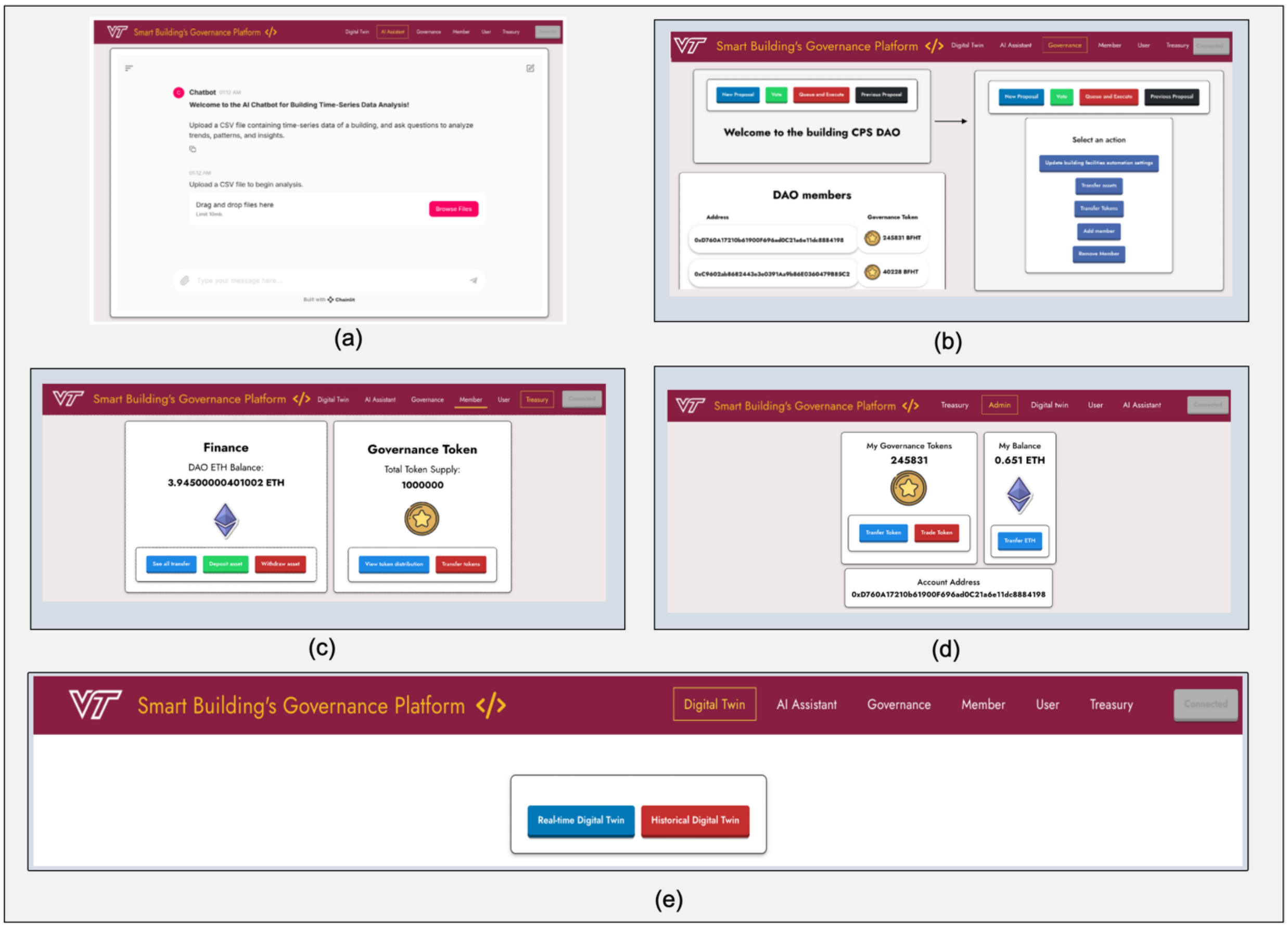

**Fig. 11** Frontend of the DA BCPS Dapp: (a) Space reservation portal (b) AI assistance portal (c) Governance Portal (d) DAO Treasury tab (e) DAO Administrator tab (f) Digital twin

## 6. Evaluation

This section outlines the evaluation and validation methodology used to assess the feasibility, usability, and usefulness of the proposed framework. A scenario-based evaluation approach was employed to simulate user interactions with the system. This validation method has been widely used in different blockchain-related studies (Tao et al. 2021, 2023; Naderi et al. 2023) and provides a feasible, effective method to demonstrate the viability of the technology in different practical contexts.  The validation process was structured around several key scenarios, including user engagement with the digital twin platform, interaction with the AI virtual assistant, and participation in the DAO governance through proposal creation and voting. Additionally, this study will evaluate the usability aspect of the proposed system using the System Usability Scale (SUS). The proposed system will also undergo qualitative assessment through expert interviews with researchers and facility managers to evaluate the platform's practical benefits and challenges for facility management applications.

### 6.1. Experiment setup

For implementation purposes, a Dapp was developed featuring three distinct stakeholders, each possessing an Ethereum account funded with 1 Sepolia testnet token. One participant deployed the DAO smart contract, including the DAO governor, governance, token, and Timelock contracts. A governance token designated as "BFHTokens" was created with a total supply of 1,000,000 units. Three accounts were each allocated 10,000 tokens, establishing their DAO membership status. Within this arrangement, one member was designated to initiate proposals, all three participated in the voting process, and one was responsible for proposal execution. In addition, user interactions with the Dapp followed the following process (Fig. 12). Steps (a)–(c) covered role assignments, account funding, and contract deployment. Steps (d) involved digital twin interactions. Steps (e)–(i) focused on DAO governance and facilities management.

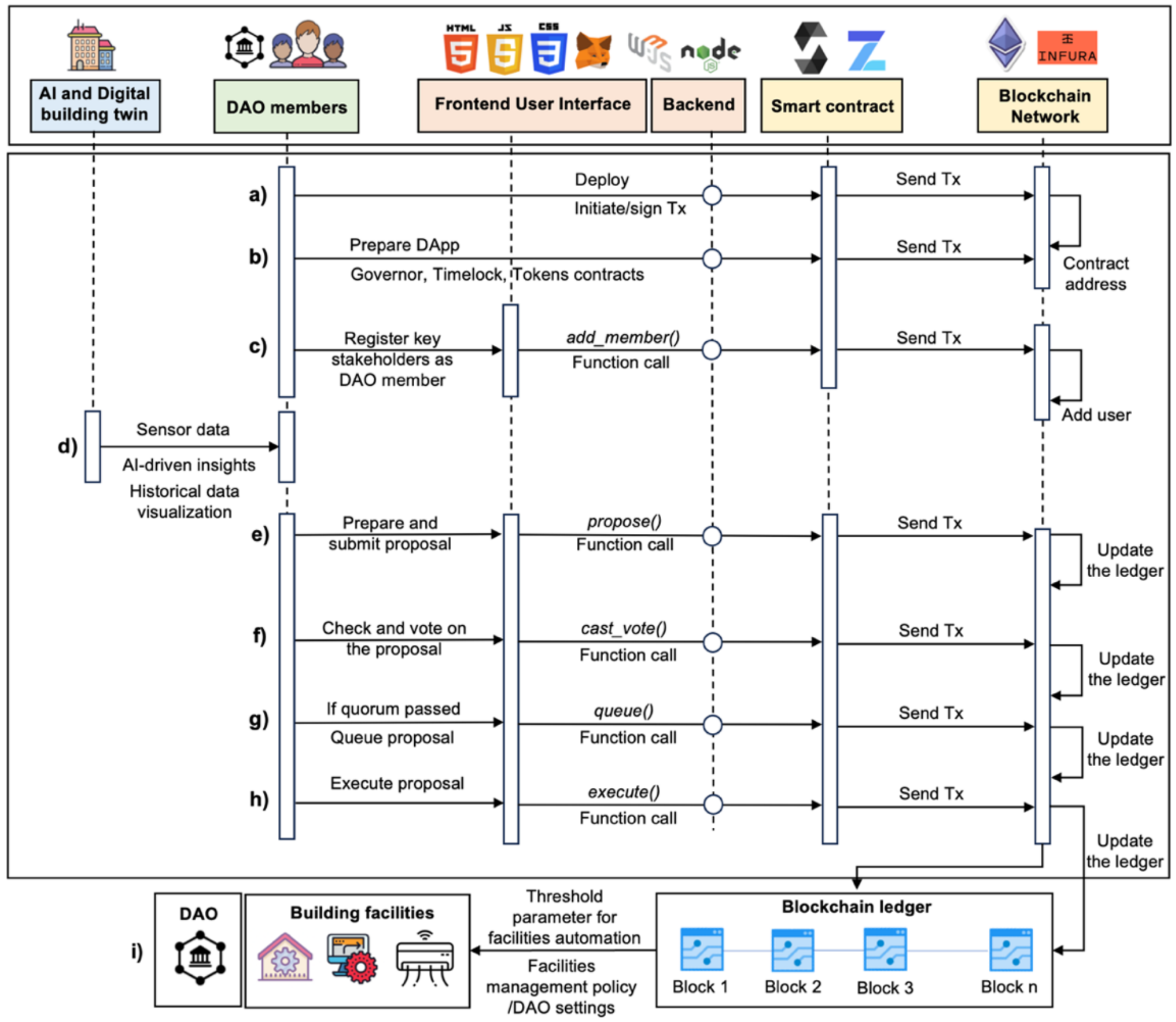

**Fig.12** Sequence diagram of implementing the Dapp

The environmental comfort parameters were configured within specific operational ranges: room temperature (20°C minimum to 27°C maximum), illuminance intensity (50 lux minimum to 150 lux maximum), relative humidity (40% minimum to 100% maximum), and carbon monoxide concentration (400 ppm minimum to 1000 ppm maximum). It is crucial to note that these parameters are used for system testing purposes and are not intended to represent optimal comfort conditions stated in established building standards such as ASHRAE Standard 55 (ASHRAE 2023) or other international comfort guidelines. The determination of optimal comfort thresholds is beyond the scope of this study. Energy usage was tracked via two smart plugs. In addition, the experimental setup incorporated several smart home devices (Fig. 13). Air quality management was facilitated through a Xiaomi Smart Air Purifier 4 Compact, which provided programmable airflow and filtration capabilities. For humidity regulation, the system integrated a Govee Smart Humidifier H7141 with adjustable output levels. Thermal comfort conditions were simulated using a Xiaomi Mi Smart Standing Fan 2, which served as a proxy for HVAC system control with variable speed settings. Illumination management was implemented through a Yeelight Smart Light Bulb W3, offering granular brightness adjustment. These specific devices were selected based on their accessible APIs and compatibility with open-source development environments, characteristics that aligned effectively with the study's research objectives and decentralized governance framework.

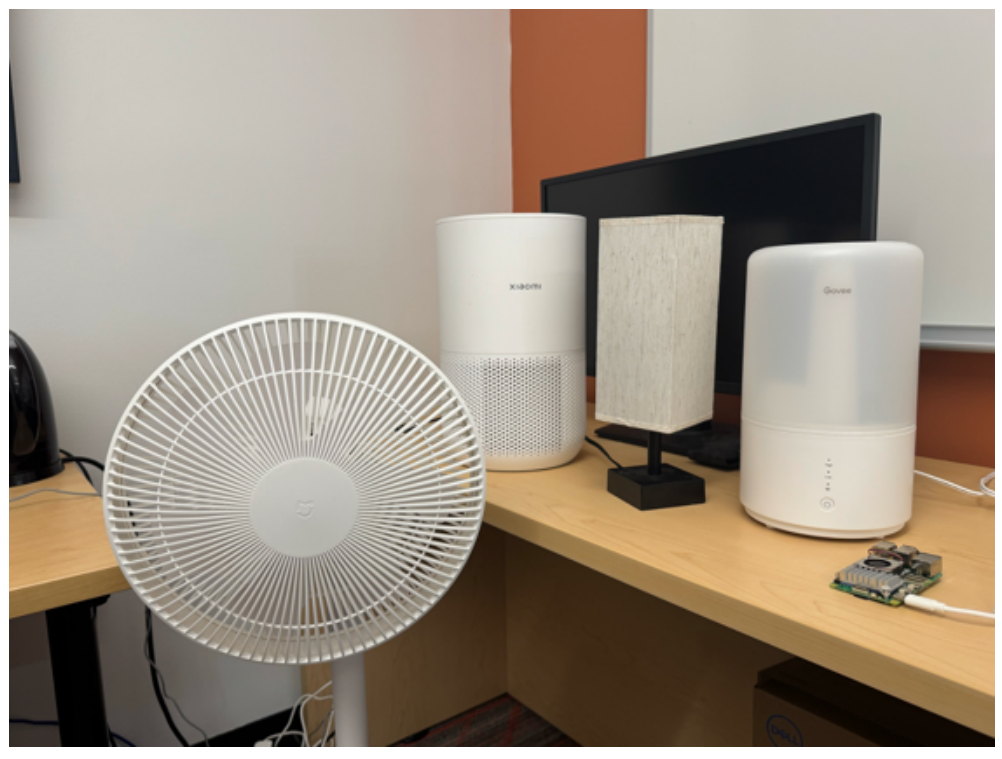

**Fig. 13** Equipment and smart home appliances used in the experiment

### 6.2. Experiment 1: User interaction with the AI assistant

This experiment aims to test the AI Virtual Assistant's ability to process building sensor data and provide actionable insights to the building manager. The experiment utilized a CSV dataset containing temperature, fan speed, occupancy levels, and energy usage over a 7-day period in a commercial building. A building manager, acting as a DAO member, uploaded the CSV file to the chatbot interface and submitted the prompt requesting data visualization and suggestions for energy optimization in the building (Fig. 14). Upon receiving a query, the AI assistant employs its reasoning capabilities to interpret the request and execute appropriate data analysis tools. The AI assistant generated multiple visualizations showing correlations between the requested parameters. Most notably, it identified that energy usage remained high during unoccupied nighttime hours, suggesting unnecessary system operation. Based on this analysis, the assistant provided several specific recommendations to the building manager: turn off fans during unoccupied hours (e.g., 10 PM-6 AM), adjust fan speed to a lower setting during occupied hours, and consider using energy-efficient lighting and HVAC systems to optimize energy usage. This experiment demonstrated the system's ability to not only visualize complex time-series data but also to derive meaningful insights that could lead to significant energy savings through decentralized governance decision-making.

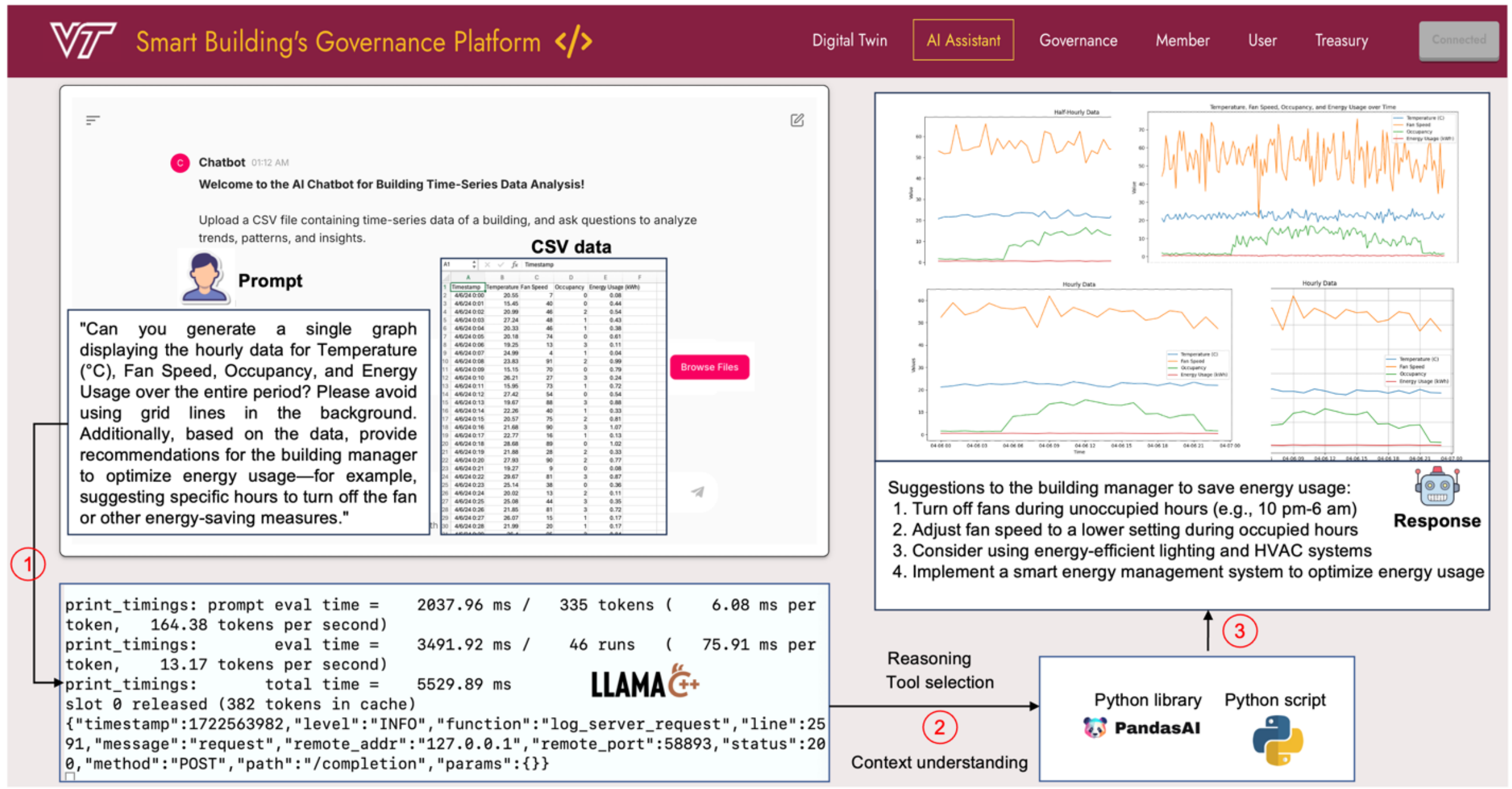

**Fig.14** Sequence diagram of implementing the Dapp

### 6.3. Experiment 2: User interaction with the Digital building twin platform

The digital twin framework provides building managers/DAO members with two complementary digital twin interfaces (Fig. 15). The first is a real-time digital twin that displays current environmental conditions, occupancy levels, and energy consumption data streamed directly from the IoT sensor network installed in Bishop-Favrao Hall. Users can navigate through different zones of the building and observe how conditions vary across spaces, enabling immediate response to anomalies or suboptimal conditions. Complementing this is the historical data digital twin, which allows users to access and visualize archived sensor data. This interface incorporates time series navigation tools that enable DAO members to examine trends, identify patterns, and compare current conditions with historical baselines. The historical twin is particularly valuable for governance decision-making as it facilitates data-driven policy development by revealing cyclical patterns in building usage, environmental fluctuations, and energy consumption over different timeframes (daily, weekly, monthly).

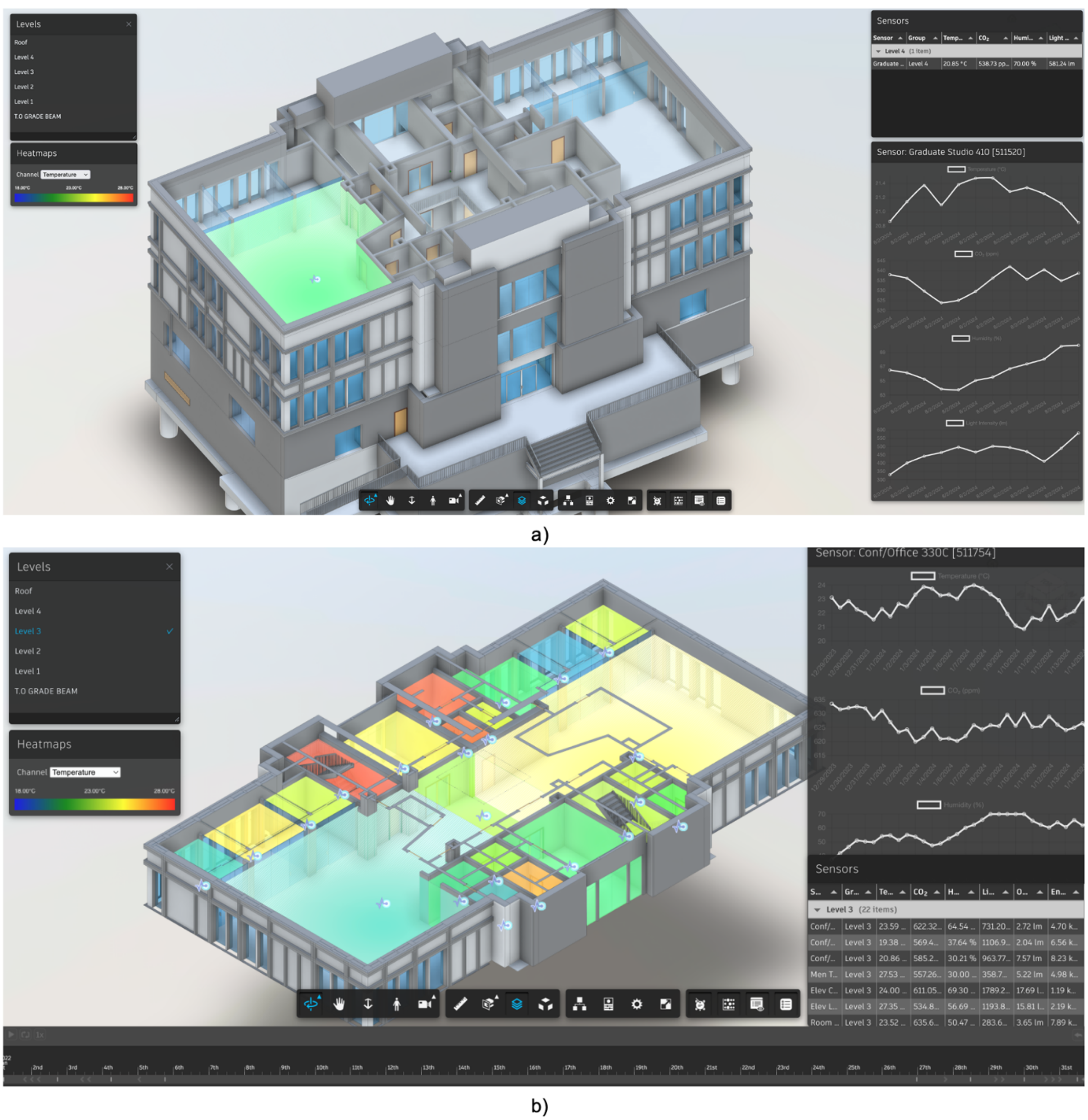

**Fig.15** Real-time digital twin b) Digital twin for historical data

### 6.4. Experiment 3: User interaction with the Decentralized governance platform

This experiment aims to assess the system's decentralized governance functionality by testing the DAO's ability to establish operational parameters for the physical space's smart devices, including modifying baseline environmental comfort thresholds such as temperature ranges, humidity levels, illuminance values, and carbon monoxide concentration limits that determine optimal occupant comfort conditions. As illustrated in Fig. 16 (steps 1 and 2), a member of the DAO submitted a proposal to alter the minimum temperature threshold to 17 degrees Celsius. Following this submission, DAO members engaged in the voting procedure (step 3), where they evaluated and cast votes either supporting or opposing the suggested modification. After successful approval through the voting process, the DAO members proceeded to queue and execute the approved proposal, which updated the threshold values in the blockchain's smart contract (step 4). The updated environmental parameters were subsequently verified as correctly recorded on the blockchain, as demonstrated in step 5.

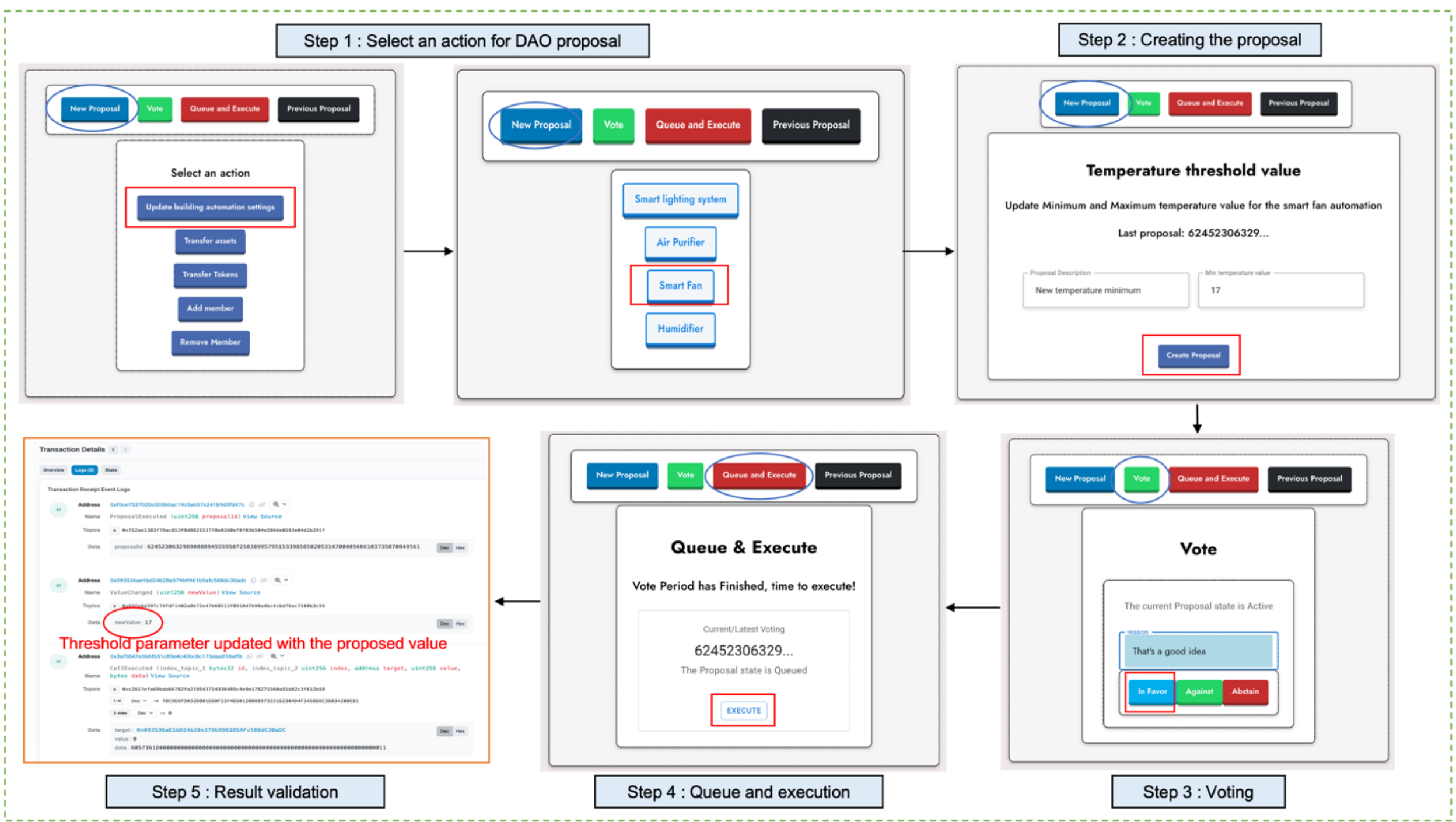

**Fig. 16** DAO governance process for setting operational thresholds for smart appliances

### 6.5. System usability evaluation

This study employed the System Usability Scale (SUS) to quantitatively assess the user-friendliness of the proposed system's key components: the decentralized governance platform, digital twin, and AI assistant. The testing involved 12 participants, which is higher than methodological approaches comparable in similar blockchain and DAO application studies (Dana et al. 2023; Peelam et al. 2025; Jeoung et al.) which typically utilized 10 participants for usability evaluation. As shown in Fig. 17 a, participants interacted with the platform by: (1) proposing and voting on building policies and automation settings; (2) analyzing IoT time series data using the AI assistant and reviewing generated visualizations; and (3) navigating the digital twin for facility management tasks. Following these interactions, participants completed a post-experiment survey containing SUS statements (Appendix A, B, and C).

### 6.6. Semi-structured interview

Semi-structured interviews were conducted with domain experts to gather comprehensive feedback on the proposed platform. This qualitative assessment provided insights into the platform's benefits and challenges regarding usability, decision-making transparency, AI assistant functionality, and the digital twin's effectiveness in facility management. There are five participants in the study. The participant count in this study is higher than in previous DAO governance research (Caviezel et al. 2023; Schmitten et al. 2023; Santos and Thesis), where typically 2-3 interviewees were consulted. The expert panel consisted of two facilities managers from Virginia Tech and three researchers with expertise in building automation, IoT data analysis, blockchain governance, and AI applications. As depicted in Fig. 17 b, interviews were conducted via Zoom, with each session lasting approximately one hour. Interview recordings were automatically transcribed by Zoom for qualitative analysis. Sample interview questions are provided in Appendix D.

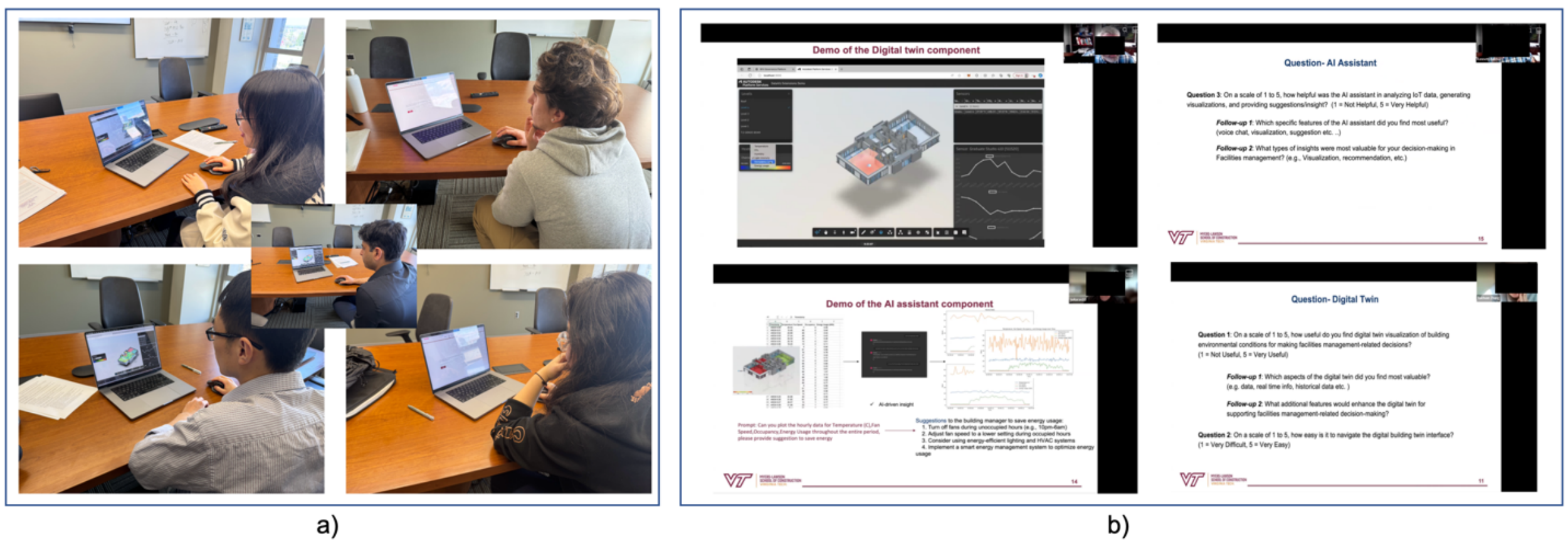

**Fig. 17** System Evaluation: (a) Usability Assessment (b) Semi-structured interview

## 7. Result and Discussion

### 7.1. Cost analysis

The Ethereum blockchain requires gas fees to process transactions, compensating for computational resources utilized on the network. These fees, denominated in Ether (ETH), can be converted to USD to demonstrate their financial impact. Gas consumption is measured in gas units, with transaction costs calculated as the product of gas consumed and gas price. The implementation of principal smart contracts like Governor, Timelock, and Tokens contracts cost approximately 0.051903 ETH (equivalent to $92.67 USD). Additional operational transactions such as DAO member registration, token transfers (both governance and Ethereum), proposal submissions, voting procedures, queuing operations, and execution processes incurred fees ranging from 0.000110 ETH ($0.20 USD) to 0.001052 ETH ($1.88 USD) per transaction. These fee calculations were performed during Sepolia testnet evaluations at an ETH rate of $1,785.36 USD (as of April 1, 2025) and are itemized in the rightmost column of Table 2.

**Table 2** The transaction cost of the proposed decentralized governance platform

| Operations | Smart contract | Gas | Transaction fee (ETH) | Transaction fee (USD) |
|---|---|---|---|---|
| Contract deployment | DAO Governor | 3,880,388 | 0.003880 | 9.15 |
| Contract deployment | Timelock controller | 1,909,795 | 0.001909 | 4.50 |
| Contract deployment | GovernanceToken | 1,971,098 | 0.001971 | 4.65 |
| Contract deployment | Facilities automation | 488,638 | 0.011985 | 28.26 |
| Contract deployment | Space reservation | 1,662,788 | 0.032158 | 75.82 |
| Adding DAO member | DAO Governor | 73,610 | 0.000110 | 0.26 |
| Proposal submission | DAO Governor | 108,168 | 0.000199 | 0.47 |
| Voting on proposal | DAO Governor | 93,186 | 0.000169 | 0.40 |
| Queuing proposal | DAO Governor | 123,769 | 0.000235 | 0.38 |
| Executing the Proposal | DAO Governor | 132,563 | 0.000238 | 0.56 |
| Governance Tokens transfer | GovernanceToken | 72,954 | 0.000139 | 0.3286 |
| Ethereum tokens transfer | Timelock controller | 21,055 | 0.001052 | 2.479 |

### 7.2. Evaluation of the AI assistant

To effectively evaluate the performance of the AI assistant, this study employs an AI as a judge evaluation framework (Son et al. 2024). This approach leverages the sophisticated understanding and generation capabilities of top-tier models such as GPT4 and Claude 3.5 to create benchmarks and assessment criteria for other open-sourced AI systems (Huang et al. 2024). Previous study has also used the framework to evaluate AI-based time-series data analysis (Merrill et al. 2024).

This framework contains three steps (Fig. 18): (step 1) generating ground truth data, (step 2) collecting predictions from our proposed AI system, and (step 3) evaluating the results against a gold standard. This systematic approach ensures a comprehensive assessment of the AI's capabilities in summarizing IoT data and providing actionable suggestions for energy efficiency and facilities management tasks.

In this initial step, we utilize ChatGPT to generate the ground truth data, serving as the gold standard for our evaluation. This process involves creating a synthetic dataset that represents various scenarios related to smart building operations. Specifically, we prompt ChatGPT to produce accurate summaries and energy-saving suggestions based on the generated IoT dataset, denoted as X_true and Y_true. In this study, the author generates 100 test sets of these data for this experiment to ensure a diverse range of scenarios. To ensure the legitimacy and coherence of the ground truth outputs, Y_true will also undergo human validation. The author will review the generated summaries and suggestions, ensuring they make sense and accurately reflect the data scenarios. Once the ground truth data is established, we proceed to collect predictions from our proposed AI assistant. This involves feeding the synthetic IoT dataset (CSV file) (X_true) into our AI model to generate its summaries and energy-saving suggestions, referred to as (Y_pred) . This step aims to evaluate how effectively our AI can interpret the data and produce meaningful insights compared to the gold standard established in Step 1. The final step in our evaluation framework is to assess the predictions made by our AI assistant against the ground truth outputs provided by ChatGPT. We employ a separate AI judge, Claude AI, to conduct this evaluation. Claude AI receives both Y_true and Y_pred and evaluates them based on predefined scoring criteria, including relevance and accuracy. The scoring system is designed to provide quantitative metrics on a scale of 0 to 100 for each criterion. Scores from multiple experiments will be aggregated to produce an overall performance score for our AI assistant. Based on the evaluation, the AI assistant achieved an average correctness score of 92% and a relevance score of 93% across 100 test cases.

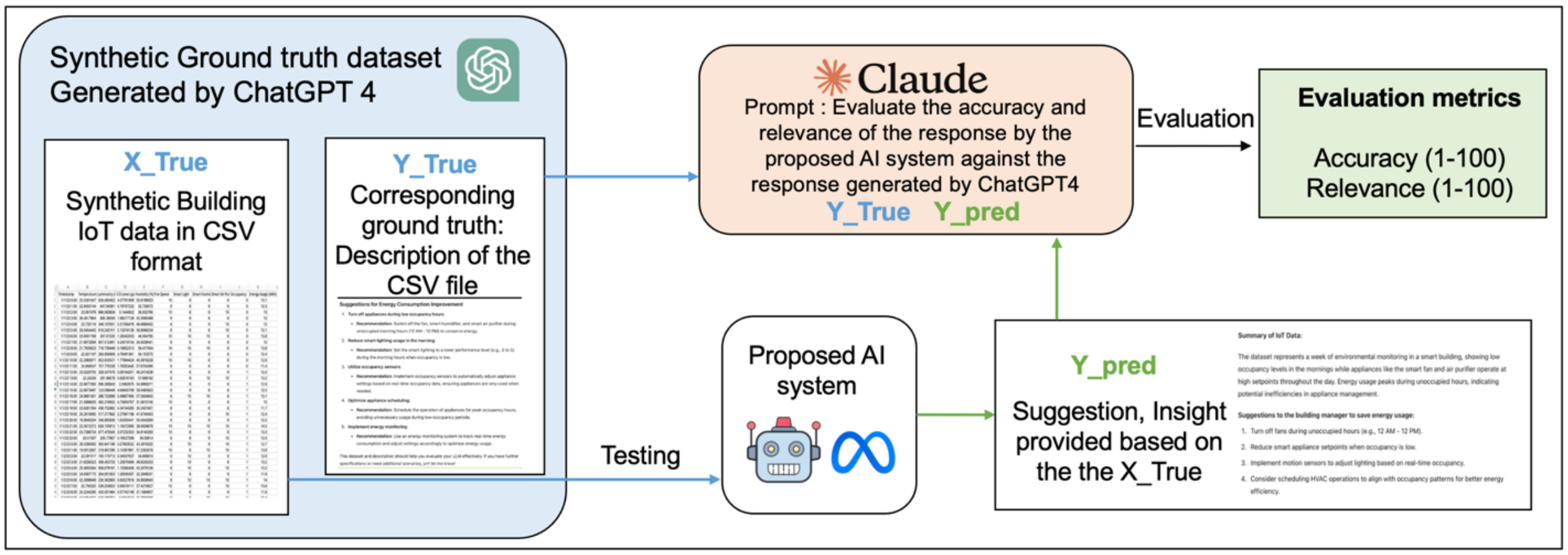

**Fig. 18** The evaluation framework of the proposed RAG system using LLM as a judge method

### 7.3. Usability evaluation

This section presents the results from the usability testing of the proposed system with 12 participants, as shown in Fig. 19 and Table 3. The digital twin-component received the highest average SUS score of 82, which according to the SUS interpretation framework by Bangor et al. (2009), falls within the "Excellent" adjective rating and corresponds to a "B" grade on the SUS grading scale. This places the digital twin in the "Acceptable" range of usability, indicating that participants found this component highly intuitive and straightforward to use.

The AI assistant component achieved an average SUS score of 79.6, which falls within the "Good" adjective rating and corresponds to a high "C" grade. This score also places the AI assistant firmly in the "Acceptable" range of the usability scale. The scores ranged from 57.5 to 100, with the majority of participants rating the system between 72.5 and 90. This moderate variability in scores may reflect the learning curve associated with formulating effective queries that were identified in the expert interviews. Nevertheless, the overall score indicates that participants found the natural language interface and data visualization capabilities to be generally user-friendly.

The decentralized governance platform received an average SUS score of 78.3, which also corresponds to a "Good" adjective rating and a high "C" grade. The governance platform scores ranged from 60 to 100, with the widest distribution among the three components, reflecting varied user experiences with the blockchain-based interface.

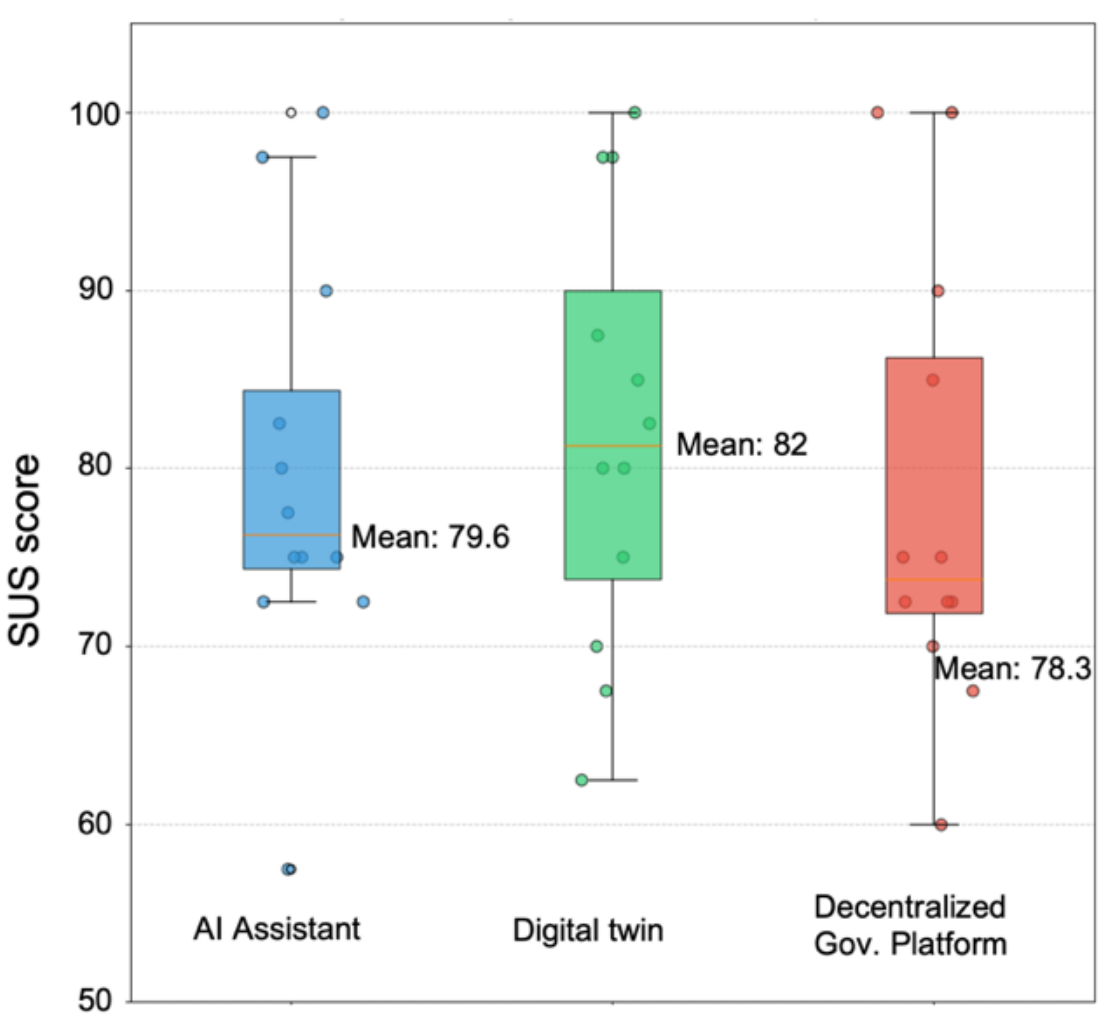

**Fig.19** SUS score of the proposed system

**Table 3** System Usability Scale (SUS) Questionnaire response

| SUS Statement (1=Strongly disagree, 2= Disagree, 3= Neutral, 4=Agree, 5 Strongly agree) | **AI assistant** | | **Digital twin** | | **Gov. Platform** | |
|---|---|---|---|---|---|---|
| | Mean | SD | Mean | SD | Mean | SD |
| 1. I think that I would like to use this system frequently. | 4.33 | 0.85 | 4.58 | 0.49 | 4.42 | 0.49 |
| 2. I found the system unnecessarily complex. | 1.91 | 0.76 | 2.25 | 1.16 | 2.08 | 0.86 |
| 3. I thought the system was easy to use. | 4.25 | 0.43 | 4.33 | 0.62 | 4.08 | 0.49 |
| 4. I think that I would need the support of a technical person to use this. | 2.25 | 1.16 | 1.83 | 0.80 | 2.25 | 0.92 |
| 5. I found the various functions in this system were well integrated. | 4.16 | 0.55 | 4.41 | 0.49 | 4.33 | 0.62 |
| 6. I thought there was too much inconsistency in this system. | 2.08 | 0.76 | 1.66 | 0.62 | 1.91 | 0.86 |
| 7. I would imagine that most people would learn to use this system quickly. | 4.41 | 0.50 | 4.33 | 0.47 | 4.16 | 0.55 |
| 8. I found the system very cumbersome to use. | 1.75 | 0.60 | 1.41 | 0.49 | 1.66 | 0.62 |
| 9. I felt very confident using the system. | 4.33 | 0.62 | 4.41 | 0.64 | 4.33 | 0.62 |
| 10. I needed to learn a lot of things before I could get going with this system. | 1.66 | 0.47 | 2.08 | 1.18 | 2.08 | 0.95 |

### 7.4. Findings from semi-structured interview

The participants' insights on perceived benefits and challenges for each system component are illustrated in Fig. 20, while their numerical ratings of usability and implementation potential are summarized in Table 4. The digital twin visualization component received the highest overall usefulness rating (4.8/5) among all system components. All five participants identified enhanced spatial visualization as a significant benefit, emphasizing the value of intuitive visual representations over traditional numerical data sheets. The real-time digital twin capabilities were particularly valued for improved situation assessment, allowing facility managers to quickly identify areas requiring attention. Historical digital twin functionality was recognized as valuable for analyzing trends and identifying patterns over time. Despite these advantages, experts highlighted several implementation challenges, with IoT infrastructure requirements being the most frequently mentioned concern. This reflects the difficulty of retrofitting existing buildings with comprehensive sensor networks, especially in older facilities with limited existing instrumentation. Additionally, passive monitoring limitations were identified, suggesting that visualization without direct control capabilities might constrain the system's utility in real-world applications.

The AI assistant component also received favorable evaluations, with an overall usefulness rating of 4.5/5. All five participants valued its natural language interface and AI-driven suggestion capabilities. Data visualization generation was also highlighted as a significant advantage for interpreting complex time-series data. However, several potential limitations were identified that could affect practical implementation. Trust and confidence in AI recommendations emerged as a notable concern, reflecting broader industry hesitation about relying on AI-generated insights without human verification. Concerns regarding Query formulation difficulties highlighted the learning curve associated with effectively communicating with AI systems through natural language. Manual IoT data upload requirements were also identified as a workflow friction point that could impede regular system use. These challenges suggest that future iterations of the AI assistant should focus on streamlining data integration and providing more transparent reasoning for recommendations to build user trust.

The decentralized governance platform received more varied assessments, with an overall usefulness rating of 3.2/5. Stakeholder inclusion and transparent decision processes were identified as key benefits by 4/5 participants, aligning with the platform's core purpose of facilitating collaborative building management. Blockchain’s immutable record-keeping capabilities were also valued for ensuring transparency and accountability in decision-making. This component faced the most significant implementation challenges among the three systems, with concerns about voting power distribution being prominent. This reflects apprehension about balancing democratic participation with appropriate weighting based on expertise and stake. Technical expertise imbalance was highlighted as a concern, pointing to the fundamental tension between inclusive decision-making and the specialized knowledge often required for facility management decisions. Transaction delays were mentioned but appeared to be a less significant concern. Notably, despite conceptual reservations, the governance platform received the highest usability/navigation rating (4.5/5), suggesting that the interface design effectively addressed user interaction needs regardless of underlying governance complexities.

Regarding implementation potential, both the digital twin and AI assistant components received identical ratings (3.8/5), indicating moderate confidence in their successful deployment. The blockchain governance platform received a lower implementation potential rating (3.2/5), reflecting the identified challenges in reconciling decentralized decision-making with technical expertise requirements. The governance platform did, however, receive a positive rating for fostering inclusive collaboration (4.0/5), confirming its alignment with the fundamental goal of community-based facility management despite implementation concerns. These expert evaluations suggest that a phased implementation approach might be most effective, beginning with the digital twin component, followed by the AI assistant, with the governance platform implemented selectively in contexts where collaborative management aligns with organizational objectives and stakeholder expertise.

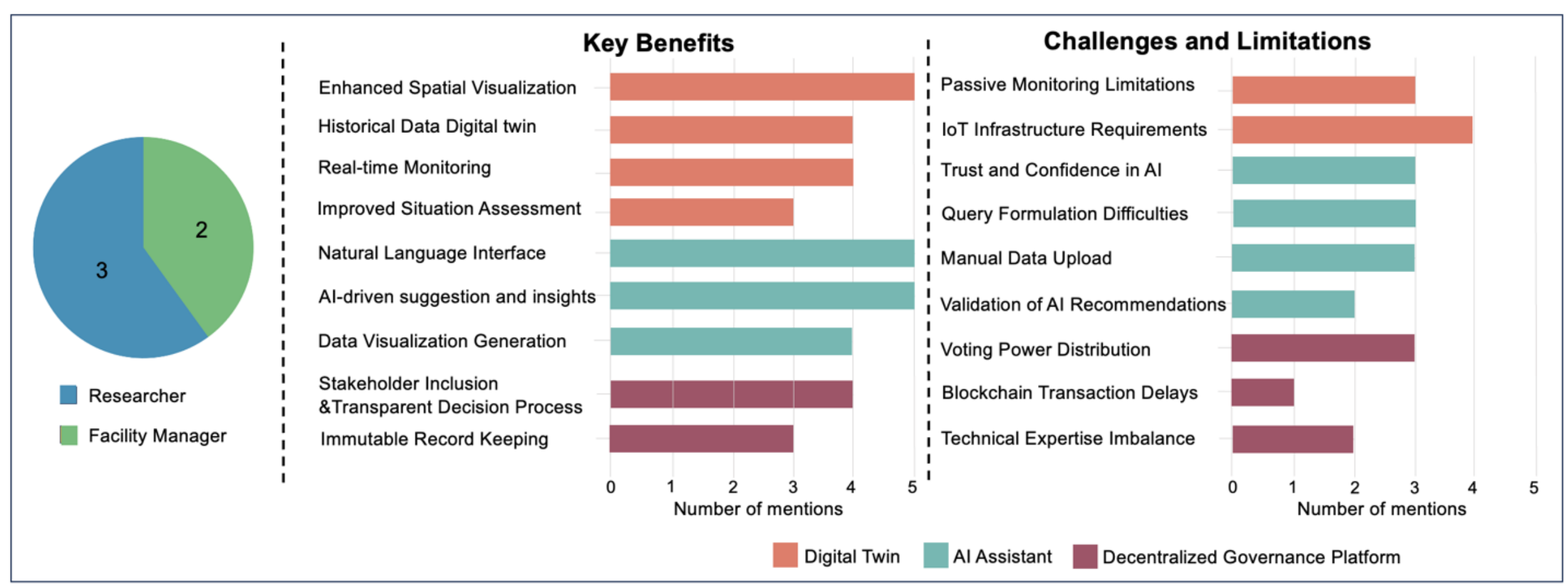

**Fig. 20** Findings from the semi-structured interview

**Table 4** Participants' ratings of the system components

| Aspect | Digital Twin | AI Assistant | Blockchain Governance |
|---|---|---|---|
| Overall Usefulness | 4.8/5 | 4.5/5 | 3.2/5 |
| Implementation Potential | 3.8/5 | 3.8/5 | 3.2/5 |
| Usability/Navigation | 4.3/5 | 4.3/5 | 4.5/5 |
| Fostering Inclusive Collaboration | N/A | N/A | 4.0/5 |

### 7.5. Scalability

The scalability of the proposed system is evaluated by assessing both its underlying blockchain infrastructure and the LLM-based AI system. For instance, Ethereum's proof-of-stake consensus mechanism imposes throughput limitations of approximately 30 transactions per second (Abrol 2022). However, in this study's experimental setup, the decentralized governance helps distribute actions over time. For instance, it is quite improbable that all DAO members will simultaneously submit proposals, vote, or execute actions, which in turn reduce the likelihood of bottlenecks.

Furthermore, the AI system's scalability evaluation focused on request handling capacity and response efficiency, measuring simultaneous request processing capabilities and response latency. In this experiment, LlamaBench is used to assess the performance of the proposed AI virtual assistant. The results indicated that execution time per command was around 5 seconds where the throughput was 33.66 tokens per second. One token is approximately equivalent to 4 English characters, and 1,500 words correspond to around 2048 tokens (OpenAI 2024).

Concurrency user request is also an important indicator of the LLM model's scalability (Yao et al. 2024). In this study, we used Llamacpp for model deployment, which allows parallelization based on the model context length. For instance, a LLAMA 3 8B model with a context window is 4096 tokens deployed on a machine with one 48G GPU can handle up to 16 concurrent requests (Fu 2024). Although the Llama 3 model we used supports a context length of up to 128k tokens, we limited it to 4096 tokens due to limited computational resources. Production deployments supporting larger user populations would benefit from expanded context length configurations and hardware platforms with increased GPU memory capacity.

### 7.6. Data Security and Privacy

Ethereum's architecture employs public-key cryptography to establish pseudonymity for users (Wang et al. 2019b). This approach ensures governance activities—including voting procedures, proposal submissions, and space reservations—are associated with pseudonymous public keys rather than personally identifiable information. While blockchain transactions remain publicly viewable, participant identities maintain anonymity and security. Furthermore, all system interactions by users, DAO members, and building occupants require transaction signing with private keys, ensuring only authorized individuals can validate and execute operations. The system's security framework also addresses transaction data accessibility. Governance proposals and voting records are deliberately transparent to enhance system integrity. Similarly, governance token holdings for each DAO member remain publicly visible, enhancing accountability within the community. This transparency model encourages participation by enabling members to verify actions and engage in governance based on verifiable information. The permanent availability of this data maintains integrity and sustains trust throughout the governance process.

### 7.7. Limitations and Future Works

This section highlights the key limitations of the proposed system and identifies opportunities for future research. One of the primary limitations of the current implementation is the reliance on Ethereum cryptocurrency for transactions within the system. The inherent volatility of Ethereum introduces financial uncertainty, making it challenging for users and DAO members to manage expenses, utility payments, and reservations effectively. This fluctuation between expected and actual costs may hinder broader adoption. Future enhancements could address this issue by integrating stablecoins such as USDT or USDC (Thanh et al. 2023), which offer more predictable and decentralized payment solutions by being pegged to reserve assets like the U.S. Dollar. Additionally, the current system depends on smart appliances and smart plugs for energy metering rather than leveraging a fully integrated smart building automation system. While this approach was sufficient for demonstrating the proof of concept, it highlights a gap that future research could address by incorporating a more comprehensive and interconnected building automation infrastructure. Another limitation of the proposed AI system is that users need to manually upload IoT data, such as CSV files, for analysis. The AI then processes these datasets to identify trends and generate insights. Future research could explore automating this process by directly linking the AI system to the IoT data repository. By enabling direct access to the database, the AI could autonomously retrieve data, analyze trends, and provide insights without requiring manual input from users.

In addition, the AI system proposed in this research is designed primarily for data analysis, trend detection, and providing recommendations based on IoT and building data. However, future research could push the boundaries by exploring AI-driven automation that not only identifies trends and insights but also autonomously implements operational adjustments on behalf of building managers. Another promising avenue for future research is the development of AI assistants that enhance human-building interaction by enabling voice- or text-based control over smart building appliances. By integrating advanced natural language processing capabilities, an AI assistant could allow occupants to control heating, lighting, ventilation, and other building systems effortlessly using simple voice commands or text inputs, further improving user experience and operational efficiency. Future research could also explore AI-driven automation for blockchain-based governance tasks. Instead of requiring DAO members to manually execute blockchain-related operations, an AI assistant could be developed to streamline these processes. Users could simply issue voice or text commands to the AI, which would then autonomously handle smart contract executions, governance decisions, and blockchain transactions on their behalf.

## 8. Conclusion

This paper presents a novel data-driven and decentralized framework for smart building facilities management, integrating web3-based governance, generative artificial intelligence, and digital building twin. The proposed framework comprises several key components. The decentralized governance platform, powered by DAO's governance, facilitates transparent decision-making and resource management. The digital twin-component provides real-time and historical visualization of environmental conditions, such as temperature, humidity, and occupancy. The large language-powered AI systems allow building managers to query important building-related data, visualization, and important insight or suggestions to improve facility management-related decision-making. The resource and code implementation for these components is available on a GitHub repository under an open-source license, allowing for further development and application of this framework beyond autonomous building management.

This study contributes to the body of knowledge in several ways: (1) Providing a novel AI-assisted, digital twin, and data-driven, blockchain-based distributed governance framework for building infrastructure facilities management. (2) Developing a full-stack, open-source DApp that can serve as a template for other AI-assisted, data-driven, and decentralized governance-related applications in other domains. (3) Demonstrating the practical implementation and evaluation of the proposed AI-assisted, data-driven, and decentralized governance framework through the prototype and case studies in an actual physical building environment. (4) Demonstrating the feasibility and benefits of large language models in enhancing human-building interaction, decision-making, and analytics of digital building twin data. (5) Offering insights into the challenges and opportunities of integrating generative AI models, such as LLMs, into decentralized governance frameworks, thereby enhancing the understanding of the interaction between AI and Web3 technologies within the built environment.

The evaluations of the system included analyses of cost efficiency, scalability of the AI and governance system, data security, and privacy.  This study also evaluates the system's usability System Usability Scale (SUS). Semi-structured interviews with researchers and facility managers were also conducted to evaluate the platform's practical benefits and challenges for facility management applications. The results from these evaluations demonstrated that the developed prototype system can potentially serve as the viable framework for future decision support systems for building managers in facilities management tasks in building infrastructure.

**Acknowledgments**
This research has received no external funding.

**Availability of data and materials**
Data will be made available on request.

**Competing interests**
The authors have no conflicts of interest to declare that are relevant to the content of this article.

**Funding**
No funding was received to assist with the preparation of this manuscript.

**Authors' contributions**
Supervision: [Alireza Shojaei]; Conceptualization: [Reachsak Ly]; Writing - original draft preparation: [Reachsak Ly]; Writing - review and editing: [Reachsak Ly]; Visualization: [Reachsak Ly]; Methodology: [Alireza Shojaei]; Project administration: [Alireza Shojaei]

**Appendix**

**Appendix A. System Usability Scale (SUS) Questionnaire for the Generative AI component of the data-driven decentralized governance system for smart building facilities management.**

| Modified SUS Statement for user experience evaluation of the Generative AI component | (1=Strongly disagree, 2= Disagree, 3= Neutral, 4=Agree, 5 Strongly agree) | | | | |
|---|---|---|---|---|---|
| | 1 | 2 | 3 | 4 | 5 |
| 1. I think I would like to use this AI assistant frequently for gaining data-driven insights for decision-making in facilities management tasks. | | | | | |
| 2. I found the AI assistant platform unnecessarily complex. | | | | | |
| 3. I thought the AI assistant platform was easy to use. | | | | | |
| 4. I think that I would need the support of a technical person to use this AI assistant. | | | | | |
| 5. I found the various functions in this AI system were well integrated. | | | | | |
| 6. I thought that there was too much inconsistency in this AI assistant. | | | | | |
| 7. I imagine that most people would learn to use this AI assistant very quickly. | | | | | |
| 8. I found the AI assistant very awkward to use. | | | | | |
| 9. I felt very confident using the AI assistant. | | | | | |
| 10. I needed to learn a lot of things before I could get going with this AI assistant. | | | | | |

**Appendix B**. **System Usability Scale (SUS) Questionnaire for the Digital Twin component of the data-driven decentralized governance system for smart building facilities management.**

| Modified SUS Statement for user experience evaluation of the Digital Twin component | (1=Strongly disagree, 2= Disagree, 3= Neutral, 4=Agree, 5 Strongly agree) | | | | |
|---|---|---|---|---|---|
| | 1 | 2 | 3 | 4 | 5 |
| 1. I think I would like to use this Digital Twin component frequently for gaining data-driven insights for decision-making in facilities management tasks. | | | | | |
| 2. I found the Digital Twin component unnecessarily complex. | | | | | |
| 3. I thought the Digital Twin component was easy to use. | | | | | |
| 4. I think that I would need the support of a technical person to use this Digital Twin component. | | | | | |
| 5. I found the various functions in this Digital Twin component were well integrated. | | | | | |
| 6. I thought that there was too much inconsistency in this Digital Twin component. | | | | | |
| 7. I imagine that most people would learn to use this Digital Twin component very quickly. | | | | | |
| 8. I found the Digital Twin component very awkward to use. | | | | | |
| 9. I felt very confident using the Digital Twin component. | | | | | |
| 10. I needed to learn a lot of things before I could get going with this Digital Twin component | | | | | |

**Appendix C**. **System Usability Scale (SUS) Questionnaire the Decentralized governance platform of the data-driven decentralized governance system for smart building facilities management.**

| Modified SUS Statement for user experience evaluation of the Decentralized governance platform | (1=Strongly disagree, 2= Disagree, 3= Neutral, 4=Agree, 5 Strongly agree) | | | | |
|---|---|---|---|---|---|
| | 1 | 2 | 3 | 4 | 5 |
| 1. I think that I would like to use this Decentralized governance platform frequently for proposing and making decision in facilities management. | | | | | |
| 2. I found the Decentralized governance platform unnecessarily complex. | | | | | |
| 3. I thought the Decentralized governance platform was easy to use. | | | | | |
| 4. I think that I would need the support of a technical person to use this Decentralized governance platform. | | | | | |
| 5. I found the various functions in this Decentralized governance platform were well integrated. | | | | | |
| 6. I thought that there was too much inconsistency in this Decentralized governance platform. | | | | | |
| 7. I imagine that most people would learn to use this Decentralized governance platform very quickly. | | | | | |
| 8. I found the Decentralized governance platform very awkward to use. | | | | | |
| 9. I felt very confident using the Decentralized governance platform. | | | | | |

| 10. I needed to learn a lot of things before I could get going with this Decentralized governance platform. | | | | | |
|---|---|---|---|---|---|

**Appendix D: Interview Guide for the Data-driven and decentralized governance platform for facilities management**

Theme 1: Usability of the platform

- Question 1: On a scale of 1 to 5, how easy is it to navigate the AI assistant interface? (1 = Very Difficult, 5 = Very Easy)
- Question 2: On a scale of 1 to 5, how easy is it to navigate the digital building twin interface? (1 = Very Difficult, 5 = Very Easy)
- Question 3: On a scale of 1 to 5, how would you rate the ease of submitting, voting on proposals, and executing the proposal via the decentralized governance platform? (1 = Very Difficult, 5 = Very Easy)
- Question 4: On a scale of 1 to 5, how intuitive do you find the overall platform's interface? (1 = Not Intuitive, 5 = Very Intuitive)
    - Follow-up 1: What specific aspects make it intuitive or non-intuitive?
    - Follow-up 2: What changes should be made, if any, to improve the usability?

Theme 2: Usefulness of the AI Assistant

- Question 5: On a scale of 1 to 5, how helpful was the AI assistant in analyzing IoT data, generating visualizations, and providing suggestions/insight? (1 = Not Helpful, 5 = Very Helpful)
    - Follow-up 1: Which specific features of the AI assistant did you find most useful?
    - Follow-up 2: What types of insights were most valuable for your decision-making? (e.g., Visualization, recommendation, etc.)
    - Follow-up 3: What additional functionality would you like to see in the AI assistant?

Theme 3: Usefulness of the Digital Twin Component

- Question 6: On a scale of 1 to 5, how useful do you find digital twin visualization of building environmental conditions for making facilities management-related decisions? (1 = Not Useful, 5 = Very Useful)
    - Follow-up 1: Which aspects of the digital twin did you find most valuable?
    - Follow-up 2: What additional features would enhance the digital twin for supporting facilities management-related decision-making?

Theme 4: Inclusivity in decision-making of the decentralized governance platform

- Question 7: On a scale of 1 to 5, how well do you think the decentralized governance platform fosters inclusivity in decision-making among different stakeholder groups for facilities management? (1 = Poorly, 5 = Very Well)
    - Follow-up 1: What specific improvements or modifications, if any, would you recommend enhancing the platform's ability to foster inclusive decision-making among all stakeholder groups?

Theme 5: Benefits and challenges

- Question 8: What do you see as the main benefits of using this platform for building facility management?
- Question 9: What are the key challenges or limitations that you foresee in implementing this platform?

Theme 6: Adoption potential

- Question 10: On a scale of 1 to 5, Please rate the following aspects of the platform (1 = Very Low, 5 = Very High)
    - The likelihood of implementing this system for future building facility management tasks.
    - The platform's effectiveness in fostering stakeholder collaboration.
    - The potential for integration with existing building management systems

**Reference**

Abrol A (2022) Solana vs Polygon vs Ethereum - Blockchain Council. https://www.blockchain-council.org/blockchain/solana-vs-polygon-vs-ethereum/. Accessed 27 Sep 2024

Adewunmi YA, Nelson M, Chigbu UE, et al (2023) A Scoping Review of Community-based Facilities Management for public services through social enterprises in developing communities. Facilities 41:868–889

Alexander K, Brown M (2006) Community-based facilities management. Facilities 24:250–268. https://doi.org/10.1108/02632770610666116

ASHRAE (2023) Standard 55 – Thermal Environmental Conditions for Human Occupancy. https://www.ashrae.org/technical-resources/bookstore/standard-55-thermal-environmental-conditions-for-human-occupancy. Accessed 22 Oct 2024

Bangor A (2009) Determining What Individual SUS Scores Mean: Adding an Adjective Rating Scale. 4:

Bujari A, Calvio A, Foschini L, et al (2021) A Digital Twin Decision Support System for the Urban Facility Management Process. Sensors 21:8460. https://doi.org/10.3390/s21248460

Cai W, Wang Z, Ernst JB, et al (2018) Decentralized applications: The blockchain-empowered software system. IEEE access 6:53019–53033

Caviezel M, Spychiger F, Stallone V (2023) Aspects for Implementations of Decentralized Autonomous Organizations (DAO) in Switzerland. In: World Conference on Information Systems and Technologies. Springer, pp 366–376

Chen K, Nadirsha TNM, Lilith N, et al (2024) Tangible digital twin with shared visualization for collaborative air traffic management operations. Transportation Research Part C: Emerging Technologies 161:104546. https://doi.org/10.1016/j.trc.2024.104546

Cheng JCP, Chen W, Chen K, Wang Q (2020) Data-driven predictive maintenance planning framework for MEP components based on BIM and IoT using machine learning algorithms. Automation in Construction 112:103087. https://doi.org/10.1016/j.autcon.2020.103087

Chotipanich S (2004) Positioning facility management. Facilities 22:364–372

Cuffe P (2018) The role of the erc-20 token standard in a financial revolution: the case of initial coin offerings. In: IEC-IEEE-KATS Academic Challenge. IEC-IEEE-KATS

Dana IR, Hidayati A, Ismail IE (2023) Blockchain Application in the Waste Trading. In: 2023 7th International Symposium on Multidisciplinary Studies and Innovative Technologies (ISMSIT). IEEE, Ankara, Turkiye, pp 1–6

Dounas T, Lombardi D (2019) Blockchain Grammars - Designing with DAOs - The blockchain as a design platform for shape grammarists' decentralised collaboration. Wellington, New Zealand, pp 293–302

Dounas T, Voeller E, Prokop S, Vele J (2022) The Architecture Decentralised Autonomous Organisation - A stigmergic exploration in architectural collaboration. Ghent, Belgium, pp 567–576

Dubey A, Jauhri A, Pandey A, et al (2024) The Llama 3 Herd of Models

Erri Pradeep AS, Yiu TW, Zou Y, Amor R (2021) Blockchain-aided information exchange records for design liability control and improved security. Automation in Construction 126:103667. https://doi.org/10.1016/j.autcon.2021.103667

Fan Y, Zhang L, Wang R, Imran MA (2023) Insight into Voting in DAOs: Conceptual Analysis and A Proposal for Evaluation Framework. IEEE Network 1–8. https://doi.org/10.1109/MNET.137.2200561

Fu Y (2024) LLM Inference Sizing and Performance Guidance. In: VMware Cloud Foundation (VCF) Blog. https://blogs.vmware.com/cloud-foundation/2024/09/25/llm-inference-sizing-and-performance-guidance/. Accessed 27 Sep 2024

Geerts GL (2011) A design science research methodology and its application to accounting information systems research. International Journal of Accounting Information Systems 12:142–151. https://doi.org/10.1016/j.accinf.2011.02.004

Gerganov G (2024) ggerganov/llama.cpp

Hong T, Wang Z, Luo X, Zhang W (2020) State-of-the-art on research and applications of machine learning in the building life cycle. Energy and Buildings 212:109831. https://doi.org/10.1016/j.enbuild.2020.109831

Huang H, Qu Y, Zhou H, et al (2024) On the Limitations of Fine-tuned Judge Models for LLM Evaluation

IFMA What is Facility Management? https://www.ifma.org/about/what-is-fm/. Accessed 27 May 2024

Jeoung J, Hong T, Jung S, et al Blockchain Framework for Occupant-centered Indoor Environment Control Using IoT Sensors

Jung Y, Kang T, Chun C (2021) Anomaly analysis on indoor office spaces for facility management using deep learning methods. Journal of Building Engineering 43:103139. https://doi.org/10.1016/j.jobe.2021.103139

Kaspersky (2021) Smart buildings threat landscape: 37.8% targeted by malicious attacks in H1 2019. In: www.kaspersky.com. https://www.kaspersky.com/about/press-releases/2019_smart-buildings-threat-landscape. Accessed 18 Nov 2023

Leaman A, Bordass B (2001) Assessing building performance in use 4: the Probe occupant surveys and their implications. Building Research & Information 29:129–143. https://doi.org/10.1080/09613210010008045

Leung M, Yu J, Yu S (2012) Investigating key components of the facilities management of residential care and attention homes. Facilities 30:611–629. https://doi.org/10.1108/02632771211270586

Ly R (2024) reachsak/decentralized-CbFM

Ly R (2025) reachsak/LLM_and_Digital_Twin_driven_Facilities_managment

Ly R, Shojaei A (2025) Decentralized autonomous organization in built environments: applications, potential and limitations. Inf Syst E-Bus Manage. https://doi.org/10.1007/s10257-025-00699-1

Ly R, Shojaei A (2024) Autonomous Building Cyber-Physical Systems Using Decentralized Autonomous Organizations, Digital Twins, and Large Language Model

Ly R, Shojaei A, Naderi H (2024) DT-DAO: Digital Twin and Blockchain-Based DAO Integration Framework for Smart Building Facility Management. In: Construction Research Congress 2024. American Society of Civil Engineers, Des Moines, Iowa, pp 796–805

m F, Lustenberger M, Martignoni J, et al (2023) Organizing projects with blockchain through a decentralized autonomous organization. Project Leadership and Society 4:100102. https://doi.org/10.1016/j.plas.2023.100102

Mascali L, Schiera DS, Eiraudo S, et al (2023) A machine learning-based Anomaly Detection Framework for building electricity consumption data. Sustainable Energy, Grids and Networks 36:101194. https://doi.org/10.1016/j.segan.2023.101194

Merrill MA, Tan M, Gupta V, et al (2024) Language Models Still Struggle to Zero-shot Reason about Time Series

Michell KA (2010) A grounded theory approach to community-based facilities management: the context of Cape Town, South Africa. University of Salford (United Kingdom)

Mutis I, Ambekar A, Joshi V (2020) Real-time space occupancy sensing and human motion analysis using deep learning for indoor air quality control. Automation in Construction 116:103237. https://doi.org/10.1016/j.autcon.2020.103237

Naderi H, Ly R, Shojaei A (2024) From Data to Value: Introducing an NFT-Powered Framework for Data Exchange of Digital Twins in the AEC Industry. In: Construction Research Congress 2024. American Society of Civil Engineers, Des Moines, Iowa, pp 299–308

Naderi H, Shojaei A, Ly R (2023) Autonomous construction safety incentive mechanism using blockchain-enabled tokens and vision-based techniques. Automation in Construction 153:104959. https://doi.org/10.1016/j.autcon.2023.104959

Nielsen SB, Sarasoja A-L, Galamba KR (2016) Sustainability in facilities management: an overview of current research. Facilities 34:535–563

Okoro CS (2023) Sustainable Facilities Management in the Built Environment: A Mixed-Method Review. Sustainability 15:3174. https://doi.org/10.3390/su15043174

OpenAI (2024) What are tokens and how to count them? https://help.openai.com/en/articles/4936856-what-are-tokens-and-how-to-count-them. Accessed 23 Sep 2024

PandasAI (2024) Sinaptik-AI/pandas-ai

Peelam MS, Kumar G, Shah K, Chamola V (2025) DEMOCRACYGUARD : Blockchain-based secure voting framework for digital democracy. Expert Systems 42:e13694. https://doi.org/10.1111/exsy.13694

Rikken O, Janssen M, Kwee Z (2023) The ins and outs of decentralized autonomous organizations (DAOs) unraveling the definitions, characteristics, and emerging developments of DAOs. Blockchain: Res Appl 100143. https://doi.org/10.1016/j.bcra.2023.100143

Saka A, Taiwo R, Saka N, et al (2024) GPT models in construction industry: Opportunities, limitations, and a use case validation. Developments in the Built Environment 17:100300. https://doi.org/10.1016/j.dibe.2023.100300

Santos F, Thesis M The DAO: A Million Dollar Lesson in Blockchain Governance

Sanzana MR, Maul T, Wong JY, et al (2022) Application of deep learning in facility management and maintenance for heating, ventilation, and air conditioning. Automation in Construction 141:104445. https://doi.org/10.1016/j.autcon.2022.104445

Schmitten J-P, Augart G, Hüsig S (2023) Decentralized Blockchain Governance and Transaction Costs in Digital Transformation: The Case of the DAO Revisited. In: 2023 Portland International Conference on Management of Engineering and Technology (PICMET). IEEE, Monterrey, Mexico, pp 1–14

Sedhom I, Khodeir LM, Fathy F (2023) Investigating current practices for achieving effective participation of stakeholders in Facilities Management. Ain Shams Engineering Journal 14:102099. https://doi.org/10.1016/j.asej.2022.102099

Seghezzi E, Locatelli M, Pellegrini L, et al (2021) Towards an Occupancy-Oriented Digital Twin for Facility Management: Test Campaign and Sensors Assessment. Applied Sciences 11:3108. https://doi.org/10.3390/app11073108

Singh M, Kim S (2019) Blockchain technology for decentralized autonomous organizations. In: Advances in Computers. Elsevier, pp 115–140

Son G, Ko H, Lee H, et al (2024) LLM-as-a-Judge & Reward Model: What They Can and Cannot Do

Tammo M, Nelson M (2014) Emergent theories for facilities management in community-based settings. Journal for Facility Management 8:22–33

Tammo M, Nelson M (2012) A critical review of the concept of facilities management in community-based contexts. Edinburgh UK

Tao X, Das M, Liu Y, Cheng JCP (2021) Distributed common data environment using blockchain and Interplanetary File System for secure BIM-based collaborative design. Automation in Construction 130:103851. https://doi.org/10.1016/j.autcon.2021.103851

Tao X, Das M, Zheng C, et al (2023) Enhancing BIM security in emergency construction projects using lightweight blockchain-as-a-service. Automation in Construction 150:104846. https://doi.org/10.1016/j.autcon.2023.104846

Thanh BN, Hong TNV, Pham H, et al (2023) Are the stabilities of stablecoins connected? Journal of Industrial and Business Economics 50:515–525

Wang S, Ding W, Li J, et al (2019a) Decentralized Autonomous Organizations: Concept, Model, and Applications. IEEE Trans Comput Soc Syst 6:870–878. https://doi.org/10.1109/TCSS.2019.2938190

Wang W, Hoang DT, Hu P, et al (2019b) A Survey on Consensus Mechanisms and Mining Strategy Management in Blockchain Networks. IEEE Access 7:22328–22370. https://doi.org/10.1109/ACCESS.2019.2896108

Xu Q, He Z, Li Z, et al (2020) An effective blockchain-based, decentralized application for smart building system management. In: Real-Time Data Analytics for Large Scale Sensor Data. Elsevier, pp 157–181

Yao Y, Jin H, Shah AD, et al (2024) ScaleLLM: A Resource-Frugal LLM Serving Framework by Optimizing End-to-End Efficiency

Zhang X, Jiang Y, Wu X, et al (2024) AIoT-enabled digital twin system for smart tunnel fire safety management. Developments in the Built Environment 18:100381. https://doi.org/10.1016/j.dibe.2024.100381

Zhao Y, Lin C-Y, Zhu K, et al (2024) Atom: Low-bit quantization for efficient and accurate llm serving. Proceedings of Machine Learning and Systems 6:196–209